\documentclass[reprint,showpacs,preprintnumbers,amssymb,nofootinbib,aps,prd]{revtex4-1}
\usepackage{epsfig,graphicx,floatflt,amssymb,rotate}
\usepackage{mathrsfs}
\usepackage{amssymb}
\usepackage{amsmath}
\usepackage{color}
\usepackage{graphicx}
\usepackage{subfig}
\usepackage{multirow}
\usepackage{pstricks}
\usepackage{float}
\usepackage[toc,page]{appendix}

\usepackage{relsize}
\def\Babar{{\mbox{\slshape B\kern-0.1em{\smaller A}\kern-0.1em B\kern-0.1em{\smaller A\kern-0.2em R}}}}

\usepackage{enumitem}

\newcommand{\ba}{\begin{array}}
\newcommand{\ea}{\end{array}}
\def\beq{\begin{equation}}
\def\eeq{\end{equation}}
\def\bea{\begin{eqnarray}}
\def\eea{\end{eqnarray}}
\def\nn{\nonumber}

\def\roughly#1{\mathrel{\raise.3ex\hbox
{$#1$\kern-.75em\lower1ex\hbox{$\sim$}}}}

\def\sla#1{\raise.15ex\hbox{$/$}\kern-.57em #1}

\def\bd{B_d^0}

\def\order{\lower 1.8ex \hbox{\LARGE\~{}}}

\def\bd0tau{B\to D \tau\nu_{\tau}}

\def\be {\begin{equation}}
\def\ee {\end{equation}}

 \usepackage[normalem]{ulem}

 \definecolor{darkgreen}{cmyk}{1,0,1,0.4}
 
 \definecolor{pink}{cmyk}{0.4,1,0.3,0}

\def\com2#1{\textcolor{red}{\it{#1}}}

\begin{document}

\title{$\mathcal{R}(D^{(*)})$ anomalies in light of Non-Minimal Universal Extra Dimension}

\author{Aritra Biswas}
\email{tbab2@iacs.res.in}
\affiliation{Department of Theoretical Physics, Indian Association for the Cultivation of Science,\\
2A $\&$ 2B Raja S.C. Mullick Road, Jadavpur, Kolkata 700 032, India}

\author{Sunando Kumar Patra}
\email{sunando.patra@gmail.com}
\affiliation{Indian Institute of Technology, North Guwahati, Guwahati 781039, Assam, India.}

\author{Avirup Shaw}
\email{avirup.cu@gmail.com}
\affiliation{Department of Theoretical Physics, Indian Association for the Cultivation of Science,\\
2A $\&$ 2B Raja S.C. Mullick Road, Jadavpur, Kolkata 700 032, India}

\begin{abstract}
We estimate contributions from  Kaluza-Klein  excitations of gauge bosons and physical charge scalar for the explanation of the lepton flavor universality violating excess in the ratios $\mathcal{R}(D)$ and $\mathcal{R}(D^*)$ in 5 dimensional Universal Extra Dimensional scenario with non-vanishing boundary localized terms. This model is conventionally known as non-minimal Universal Extra Dimensional model. We obtain the allowed parameter space in accordance with constraints coming from $B_c \to \tau \nu$ decay, as well as those from the electroweak precision tests. 
\end{abstract}

\maketitle


\vspace*{-0.1cm} 
\section{Introduction}
The Standard model (SM) is till date the most successful model in explaining our understanding of the fundamental particles that are the building blocks of nature. Its predictive capability as well as robustness have rigorously and repeatedly been put to test over the last fifty years, with the final piece of the puzzle, the Higgs' boson, being discovered back in 2012~\cite{atlas, cms}. However, we do know that this model is still not the complete picture, as there exist experimental signatures for the presence of new physics (NP), such as massive neutrinos, Dark Matter (DM) etc., that cannot be accounted for by this model. Hence, the phenomenology of particle physics at this point of time is not only subject to precision tests for the SM, but also on the lookout for observables that show a deviation from their SM predictions. These observables can then be used as a probe for the favorable kind among the various NP models that exist in the literature. The significance of the deviation of these observables from SM can be used to rule out or constrain these available models, or even predict the type of NP that one can hope to probe.

To this end, heavy flavor physics has emerged as a powerful tool over the past three decades. Tensions between SM expectations and experimental results have been found for observables such as the isospin asymmetry $A_I(B\rightarrow K\mu^+\mu^-)$~\cite{Aaij:2012cq}, the longitudinal polarization fraction in $B_s\rightarrow K^*K^*$~\cite{Aaij:2011aj, DescotesGenon:2011pb}, $\mathcal{R}(D^*)=\frac{\text{BR}(B\rightarrow D^*\tau\nu_{\tau})}{\text{BR}(B\rightarrow D^*l\nu_l)}$, $\mathcal{R}(D)=\frac{\text{BR}(B\rightarrow D\tau\nu_{\tau})}{\text{BR}(B\rightarrow Dl\nu_l)}$~\cite{Kamenik:2008tj}-\cite{Abdesselam:2016cgx} and $\mathcal{R}_K=\frac{\text{BR}(B\rightarrow K\mu^+\mu^-)}{\text{BR}(B\rightarrow Ke^+e^-)}$~\cite{Bobeth:2007dw, Aaij:2014ora}.
Among these, the $\mathcal{R}(D)$ and $\mathcal{R}(D^*)$ are particularly interesting because $B\to D^{(*)} \tau \nu_{\tau}$ are tree level processes in the SM and as $|V_{cb}|$ and some form factor parameters get canceled in the ratios, they are theoretically cleaner. The latest world averages of the experimental results of these ratios, combined (table \ref{tab:RDRDst}), are about $4\sigma$ away from the latest SM
predictions for these ratios. These ratios thus point to a tantalizing prospect of beyond SM physics.  
In this paper we aim at exploring these ratios in the Universal Extra Dimensional (UED) model~\cite{acd}. 

The scenario of UED has been built up with one extra space-like flat dimension ($y$), compactified on a circle $S^1$ of radius $R$, which is accessed by all the SM particles. Eventually from the 4 dimensional (4D) point of view, this model possesses SM particles along with their infinite number of Kaluza-Klein (KK)-partners specified by the so called KK-number ($n$). This number represents the discretized momentum in the direction of the extra dimension. One imposes an extra $Z_2$ symmetry ($y \leftrightarrow -y$) to generate the chiral fermions in the theory. We therefore eventually come up with two special points $y =0$ and $y = \pi R$ called the {\em fixed points} along the $y$ direction and also with a remnant symmetry called KK-parity $(-1)^n$. This symmetry has several consequences. For example conservation of this KK-parity ensures that the lightest Kaluza-Klein particle (LKP) with KK-number being unity ($n = 1$) cannot decay to a pair of SM particles and is absolutely stable. Consequently the LKP can be considered as a potential DM candidate for this scenario \cite{ued_dm, relic}. Moreover, this model can address other longstanding unsolved issues due to the SM, like gauge coupling unification \cite{ued_uni}, neutrino mass \cite{ued_nutrino}, fermion mass hierarchy \cite{hamed} etc.

The mass of the $n^{th}$ KK-partner of any SM particle is given by $\sqrt{(m^2+(nR^{-1})^2)}$ where $m$ is the zero-mode mass (SM particle mass) which is small compared to $R^{-1}$. Hence, this scenario suffers from almost degenerate particle spectrum
at each KK-level. However, the corresponding radiative corrections in 5 dimensions (5D) offer a remedy to this situation~\cite{rad_cor_georgi, rad_cor_cheng}. There are two kinds of radiative corrections, the bulk corrections which are finite and nonzero for KK excitations of gauge bosons only, and the boundary localized corrections (depending logarithmically on the cut-off\footnote{Being an extra dimensional theory UED can be considered as an effective theory characterized by a cut-off scale $\Lambda$.} scale $\Lambda$\cite{rad_cor_georgi}) which are embedded as contributions to the 4D Lagrangian located at the the two fixed points of the orbifold. These boundary terms behave as counter-terms for cut-off dependent loop-induced contributions. The minimal version of UED assumes that one could tune the boundary terms in a way so that 5D radiative corrections exactly vanish at the cutoff scale $\Lambda$. However, this assumption can be discarded and instead of actually estimating the boundary localized corrections one might consider kinetic, mass as well as other interaction terms to parametrize these unknown corrections. This model is collectively called non-minimal UED (NMUED). Coefficients of the several boundary localized terms (BLTs) along with the radius of compactification ($R$) can be viewed as free parameters of this model and one can constrain these parameters using different experimental data. One can find various phenomenological analyses in the framework of NMUED from different perspectives in the literature. For example, bounds on the values of the coefficients of the boundary localized terms are obtained from the consideration of electroweak observables \cite{flacke}, $S$, $T$ and $U$ parameters \cite{delAguila_STU, flacke_STU}, relic density \cite{tommy, ddrs2}, production and decay of the SM Higgs boson \cite{tirtha}, study of LHC experiments \cite{asesh_lhc1, lhc}, $R_b$ \cite{zbb}, branching ratio of  $B_s \rightarrow \mu^+ \mu^-$ \cite{bmm} and $B \rightarrow X_s \gamma$ \cite{bsg}, flavour changing rare top decay \cite{Dey:2016cve} and unitarity of scattering amplitudes involving KK-excitations \cite{Jha:2016sre}.

In this article we explore the effects of the parameters of this NMUED model on the $\mathcal{R}(D)$ and $\mathcal{R}(D^*)$ observables. One should note here that this exercise will not be possible in the MUED model. This is because in MUED, the orthogonality relations between the KK-wave functions of different fields prohibit tree level couplings between a pair of SM fermions and the KK-partner of gauge bosons and charged Higgs. However, the effects of non-minimality allow one to generate these couplings specified by non-zero even KK-number(s). 
Considering contributions from the (potentially infinite) gauge bosons alone will not affect the ratios $\mathcal{R}(D^{(*)})$ in any way 
\footnote{This contribution will however change the binned $B\rightarrow D^*\ell\nu_{\ell}$ scenario, but that is not of immediate interest in our present article.}. However, on considering the contribution coming from the (large number of) possible Higgs scalars, one encounters a lepton-flavor dependent coefficient.

We organize the article in the following way: we introduce the NMUED model in section \ref{NMUED}, express $\mathcal{R}(D^{(*)})$ in terms of the model parameters in section \ref{rdrdst_formalism}, describe the present experimental status of the ratios and glean information about the model parameters from experimental results ($\mathcal{R}(D^{(*)})$ and $B_c\rightarrow\tau\nu_{\tau}$) in sections \ref{chwfit} to \ref{modelparams}. In section \ref{ewpt}, we put additional constraints on NMUED parameters from the experimental fit results of oblique electroweak precision parameters.

\section{KK-parity conserving NMUED scenario in a nutshell}\label{NMUED} 

Here we briefly discuss the NMUED model and the parameters therein that are relevant for the present analysis. For a detailed discussion we refer to \cite{asesh_lhc1}, \cite{zbb, bmm}, \cite{Dvali}-\cite{ddrs1}. In this scenario we do maintain the boundary terms to be equal\footnote{One can proceed with unequal strengths for the boundary terms. In that case the KK parity will not be restored as a result of non-conservation. A detailed discussion on the phenomenology in such KK parity non-conserving cases can be found in~\cite{ddrs1, lhc, Bhattacherjee:2008kx}.} at both boundary points ($y=0$ and $y=\pi R$). This will preserve a discrete ${Z}_2$ symmetry which exchanges $y \longleftrightarrow (y- \pi R)$, hence the KK parity is restored in this scenario and makes the LKP stable. Eventually one has the potential DM candidate (e.g., first excited KK-state of photon) in this scenario. One can find an extensive work on DM relic density and related issues in this NMUED model in \cite{ddrs2}.
 
We start with the action for 5D fermionic fields including their boundary localised kinetic term (BLKT) of strength $r_f$ (\cite{schwinn, ddrs2, bmm,bsg}):
\vspace*{-0.1cm}
\begin{eqnarray} 
S_{fermion} &=& \int d^5x \bigg[ \bar{\Psi}_L i \Gamma^M D_M \Psi_L \nonumber \\ 
&+& r_f\{\delta(y)+\delta(y - \pi R)\} \bar{\Psi}_L i \gamma^\mu D_\mu P_L\Psi_L  
\nonumber \\
&+& \bar{\Psi}_R i \Gamma^M D_M \Psi_R \nonumber \\
&+& r_f\{\delta(y)+\delta(y - \pi R)\}\bar{\Psi}_R i \gamma^\mu D_\mu P_R\Psi_R \bigg],\nonumber\\
\label{factn}
\end{eqnarray}

where $\Psi_L(x,y)$ and $\Psi_R(x,y)$ are the 5D four component Dirac spinors, which can be expressed in terms of two component spinors \cite{schwinn, ddrs2, bmm, bsg}:

\begin{equation} 
\Psi_L(x,y) = \begin{pmatrix}\phi_L(x,y) \\ \chi_L(x,y)\end{pmatrix}
=   \sum_n \begin{pmatrix}\phi^{(n)}_L(x) f_L^n(y) \\ \chi^{(n)}_L(x) g_L^n(y)\end{pmatrix}, 
\label{fermionexpnsn1}
\end{equation}
\begin{equation} 
\Psi_R(x,y) = \begin{pmatrix}\phi_R(x,y) \\ \chi_R(x,y) \end{pmatrix} 
=   \sum_n \begin{pmatrix}\phi^{(n)}_R(x) f_R^n(y) \\ \chi^{(n)}_R(x) g_R^n(y) \end{pmatrix}. 
\label{fermionexpnsn2} 
\end{equation}\\

Here $f_{L(R)}$ and $g_{L(R)}$ are the KK-wave functions that can be written in the following form~\cite{carena, flacke, ddrs2, bmm, bsg}: 
\begin{eqnarray}
f_L^n = g_R^n = N^f_n \left\{ \begin{array}{rl}
                \displaystyle \frac{\cos\left[m_{f^{(n)}} \left (y - \frac{\pi R}{2}\right)\right]}{\cos[ \frac{m_{f^{(n)}} \pi R}{2}]}  &\mbox{for $n$ even,}\\
                \displaystyle \frac{{-}\sin\left[m_{f^{(n)}} \left (y - \frac{\pi R}{2}\right)\right]}{\sin[ \frac{m_{f^{(n)}} \pi R}{2}]} &\mbox{for $n$ odd,}
                \end{array} \right.
                \label{flgr}
\end{eqnarray}
and
\begin{eqnarray}
g_L^n =-f_R^n = N^f_n \left\{ \begin{array}{rl}
                \displaystyle \frac{\sin\left[m_{f^{(n)}} \left (y - \frac{\pi R}{2}\right)\right]}{\cos[ \frac{m_{f^{(n)}} \pi R}{2}]}  &\mbox{for $n$ even,}\\
                \displaystyle \frac{\cos\left[m_{f^{(n)}} \left (y - \frac{\pi R}{2}\right)\right]}{\sin[ \frac{m_{f^{(n)}} \pi R}{2}]} &\mbox{for $n$ odd.}
                \end{array} \right.
\end{eqnarray}
$N^f_n$ is the normalisation constant for $n^{th}$ KK-mode which can be readily derived from orthonormality conditions \cite{ddrs2, bmm, bsg}:
\begin{equation}\label{orthonorm}
\begin{aligned}
&\left.\begin{array}{r}
                  \int_0 ^{\pi R}
dy \; \left[1 + r_{f}\{ \delta(y) + \delta(y - \pi R)\}\right]f_L^mf_L^n\\
                  \int_0 ^{\pi R}
dy \; \left[1 + r_{f}\{ \delta(y) + \delta(y - \pi R)\}\right]g_R^mg_R^n
\end{array}\right\}=\delta^{n m}~;\\
&\left.\begin{array}{l}
                 \int_0 ^{\pi R}
dy \; f_R^mf_R^n\\
                 \int_0 ^{\pi R}
dy \; g_L^mg_L^n
\end{array}\right\}=\delta^{n m}~,
\end{aligned}
\end{equation}

and it is given by:
{\small
\vspace*{-.3cm}
\begin{equation}\label{norm}
N^f_n=\sqrt{\frac{2}{\pi R}}\Bigg[ \frac{1}{\sqrt{1 + \frac{r^2_f m^2_{f^{(n)}}}{4} + \frac{r_f}{\pi R}}}\Bigg].
\end{equation}
}

The mass of the $n^{th}$ KK-excitation ($m_{f^{(n)}}$) satisfies the following transcendental equations \cite{carena, ddrs2, bmm,bsg}: 
\begin{eqnarray}
  \frac{r_{f} m_{f^{(n)}}}{2}= \left\{ \begin{array}{rl}
         -\tan \left(\frac{m_{f^{(n)}}\pi R}{2}\right) &\mbox{for $n$ even,}\\
          \cot \left(\frac{m_{f^{(n)}}\pi R}{2}\right) &\mbox{for $n$ odd.}
          \end{array} \right.   
          \label{fermion_mass}      
 \end{eqnarray}


The action for the Yukawa interaction with the corresponding boundary localised terms of strength $r_y$ is given by \cite{bmm,bsg}:  

\begin{eqnarray}
\label{yukawa}
S_{Yukawa} &=& -\int d^5 x  \Big[\lambda^5_t\;\bar{\Psi}_L\widetilde{\Phi}\Psi_R \nonumber\\ 
  &+&r_y \;\{ \delta(y) + \delta(y-\pi R) \}
  \lambda^5_t\bar{\phi_L}\widetilde{\Phi}\chi_R+\textrm{h.c.}\Big].\nonumber\\
\end{eqnarray}

Here $\lambda^5_t$ represents the 5D coupling for the Yukawa interaction for the third generation. 
Inserting the KK-expansions for fermions (eqs.~\ref{fermionexpnsn1} and \ref{fermionexpnsn2})
in the actions (eq.~\ref{factn} and eq.~\ref{yukawa}) one obtains the bi-linear terms involving the doublet and singlet representations of the quarks. The mass matrix for the $n^{th}$ KK-level is~\cite{bmm,bsg}: 

\begin{align}
\label{fermion_mix}
-\begin{pmatrix}
\bar{\phi_L}^{(n)} & \bar{\phi_R}^{(n)}
\end{pmatrix}
\begin{pmatrix}
m_{f^{(n)}}\delta^{nm} & m_{t} {\mathscr{I}}^{nm}_1 \\ m_{t} {\mathscr{I}}^{mn}_2& -m_{f^{(n)}}\delta^{mn}
\end{pmatrix}
\begin{pmatrix}
\chi^{(m)}_L \\ \chi^{(m)}_R
\end{pmatrix}\nonumber\\
+{\rm h.c.}.
\end{align}
Here, $m_t$ stands for the SM top quark mass and $m_{f^{(n)}}$ are the solutions of transcendental equations given in eq.~\ref{fermion_mass}. The overlap integrals (${\mathscr{I}}^{nm}_1$ and ${\mathscr{I}}^{nm}_2$) are given by \cite{bmm,bsg}:
\begin{eqnarray}
{\mathscr{I}}^{nm}_1&=&\left(\frac{1+\frac{r_f}{\pi R}}{1+\frac{r_y}{\pi R}}\right)\times\nonumber\\&& \int_0 ^{\pi R}\hspace*{-0.5cm}dy
\left[ 1+ r_y \{\delta(y) + \delta(y - \pi R)\} \right] g_{R}^m f_{L}^n~,  \\ {\mathscr{I}}^{nm}_2&=&\left(\frac{1+\frac{r_f}{\pi R}}{1+\frac{r_y}{\pi R}}\right)\times\int_0 ^{\pi R}\hspace*{-0.5cm}dy~
 g_{L}^m f_{R}^n~.
\end{eqnarray}
The integral ${\mathscr{I}}^{nm}_1$ is non zero for both the cases of $n=m$ and $n\neq m$. However for $r_y = r_f$, this integral equals 1 (when $n =m$) or 0 (for $n \neq m$). The integral ${\mathscr{I}}^{nm}_2$ is non vanishing only when  $n=m$ and equal to 1 in the limit $r_y = r_f$. To avoid the complicacy of mode mixing and construct a simpler form of fermion mixing matrix  we choose an equality condition ($r_y$=$r_f$) in our analysis~\cite{zbb,bmm,bsg}. With this motivation, we will keep this equality ($r_y=r_f$) in the rest of our analysis \footnote{However, in general one can proceed with unequal strengths of boundary terms for Yukawa and kinetic interaction for fermions.}. 

Applying the above equality criteria, the resulting mass matrix (given in eq.~\ref{fermion_mix}) can be diagonalised by following bi-unitary transformations for the left- and right-handed fields respectively \cite{bmm,bsg}:
\begin{eqnarray}
U_{L}^{(n)}=\begin{pmatrix}
\cos\alpha_{tn} & \sin\alpha_{tn} \\ -\sin\alpha_{tn} & \cos\alpha_{tn}
\end{pmatrix},\\U_{R}^{(n)}=\begin{pmatrix}
\cos\alpha_{tn} & \sin\alpha_{tn} \\ \sin\alpha_{tn} & -\cos\alpha_{tn}
\end{pmatrix},
\end{eqnarray}
where 
\begin{equation}
\alpha_{tn}=\frac12\tan^{-1}\left(\frac{m_{t}}{m_{f^{(n)}}}\right)
\label{alphatn}
\end{equation}
is the mixing angle. The mass ($T^1_t$ and $T^2_t$) and gauge ($\Psi_L(x,y)$ and $\Psi_R(x,y)$) eigenstates are related by the following relations \cite{bmm,bsg}:
%
\begin{tabular}{p{8cm}p{8cm}}
{\begin{align}
&{\phi^{(n)}_L} =  \cos\alpha_{tn}T^{1(n)}_{tL}-\sin\alpha_{tn}T^{2(n)}_{tL}, \\
&{\chi^{(n)}_L} =  \cos\alpha_{tn}T^{1(n)}_{tR}+\sin\alpha_{tn}T^{2(n)}_{tR},
\end{align}}
{\begin{align}
&{\phi^{(n)}_R} =  \sin\alpha_{tn}T^{1(n)}_{tL}+\cos\alpha_{tn}T^{2(n)}_{tL}, \\
&{\chi^{(n)}_R} =  \sin\alpha_{tn}T^{1(n)}_{tR}-\cos\alpha_{tn}T^{2(n)}_{tR}.
\end{align}}
\end{tabular}
%
The mass eigenvalue at the $n^{th}$ KK-level is $M_{t^{(n)}} \equiv \sqrt{m_{t}^{2}+m^2_{f^{(n)}}}$ and is the same for both physical eigenstates $T^{1(n)}_t$ and $T^{2(n)}_t$. 

Let us concentrate on the free action (governed by $SU(2)_L \times U(1)_Y$ gauge group) of 5D gauge and scalar fields with their respective BLKTs \cite{flacke, zbb, bmm, bsg, Dey:2016cve, gf}:

\begin{eqnarray}
S_{gauge} &=& -\frac{1}{4}\int d^5x \bigg[ W^i_{MN} W^{iMN}\nonumber \\
&+& r_W \left\{ \delta(y) +  \delta(y - \pi R)\right\} W^i_{\mu\nu} W^{i\mu \nu}\nonumber \\
&+& B_{MN} B^{MN}\nonumber \\
&+& r_B \left\{ \delta(y) +  \delta(y - \pi R)\right\} B_{\mu\nu} B^{\mu \nu}\bigg],
\label{pure-gauge}
\end{eqnarray}

\begin{eqnarray} 
S_{scalar} &=& \int d^5x \bigg[ (D_{M}\Phi)^\dagger(D^{M}\Phi) \nonumber \\
&+& r_\phi \left\{ \delta(y) +  \delta(y - \pi R)\right\} (D_{\mu}\Phi)^\dagger(D^{\mu}\Phi) \bigg].
\label{higgs}
\end{eqnarray}

Here, $r_W$, $r_B$ and $r_\phi$ parametrise the strength of the BLKTs for the respective fields. 5D field strength tensors are given below:
\begin{eqnarray}\label{ugfs}
W_{MN}^i &\equiv& (\partial_M W_N^i - \partial_N W_M^i-{\tilde{g}_2}\epsilon^{ijk}W_M^jW_N^k),\\ \nonumber
B_{MN}&\equiv& (\partial_M B_N - \partial_N B_M).
\end{eqnarray}
$W^i_M (\equiv W^i_\mu, W^i_4)$ and $B_M (\equiv B_\mu, B_4)$ ($M=0,1 \ldots 4$) are the 5D gauge fields corresponding to $SU(2)_L$ and $U(1)_Y$ gauge groups respectively. 5D covariant derivative is defined as $D_M\equiv\partial_M+i{\tilde{g}_2}\frac{\sigma^{i}}{2}W_M^{i}+i{\tilde{g}_1}\frac{Y}{2}B_M$, with the 5D gauge coupling constants ${\tilde{g}_2}$ and ${\tilde{g}_1}$. ${\sigma^{i}}\over 2$ and $\frac Y2$ are the corresponding generators of the gauge groups. $i$ is the $SU(2)_L$ group index and runs from 1 to 3. $\Phi=\left(\begin{array}{cc}
\phi^+\\\phi^0\end{array}\right)$ is the 5D Higgs doublet.
Appropriate KK-expansion of the gauge and scalar fields which are involved in the above actions (eqs.\;\ref{pure-gauge} and \ref{higgs}) can be schematically written as \cite{bmm, bsg, gf}:
\begin{equation}\label{Amu}
V_{\mu}(x,y)=\sum_n V_{\mu}^{(n)}(x) a^n(y)
\end{equation}
\begin{equation}\label{A4}
V_{4}(x,y)=\sum_n V_{4}^{(n)}(x) b^n(y),
\end{equation}
and
\begin{equation}\label{chi}
\Phi(x,y)=\sum_n \Phi^{(n)}(x) h^n(y).
\end{equation}
In the above $V$ generically represents both the $SU(2)_L$ and $U(1)_Y$ gauge bosons.
 
Before going further, we would like to make some clarifying remarks which could help the reader understand the following field and the corresponding KK-wave function structure of the gauge as well as the scalar particles. First of all, KK-decomposition of neutral gauge bosons become very complicated due to the fact that $B$ and $W^3$ mix in the bulk as well as on the boundary. So, unless $r_W=r_B$, it would not be possible to diagonalise the bulk and boundary actions simultaneously by the same 5D field redefinition\footnote{However, in general one can deal with $r_W\neq r_B$, but in this case the mixing term between $B$ and $W^3$ in the bulk and on the boundary points generate off-diagonal terms in the neutral gauge boson mass matrix.}. In the following we will stick to the $r_W=r_B$ equality condition \cite{zbb, bmm, bsg, Dey:2016cve, gf}. As a consequence, in this case one has the same structure of mixing between KK-excitations of the neutral component of the gauge fields (i.e., the mixing between $W^{3(n)}$ and $B^{(n)}$) as the MUED scenario. Eventually, the mixing between $W^{3(1)}$ and $B^{(1)}$ (i.e., the mixing at the first KK-level) gives the $Z^{(1)}$ and $\gamma^{(1)}$. This $\gamma^{(1)}$ is absolutely stable which can not decay to pair of SM particles by the conservation KK-parity and possesses the lowest mass among the first excited KK states in the NMUED particle spectrum. Therefore, this $\gamma^{(1)}$ has been treated as the DM candidate of this scenario \cite{ddrs2}. From now and onwards we use $r_V$ as the generic BLKT parameter for gauge bosons.  

Eqs.\;\ref{pure-gauge} and \ref{higgs} must be supplemented by the gauge-fixing action.  In the following, we have considered the following gauge fixing action appropriate for NMUED model \cite{zbb, bmm, bsg, Dey:2016cve, gf}. A detailed study on gauge fixing action/mechanism in NMUED can be found in ref.\;\cite{gf}. 
\begin{eqnarray} 
S_{GF} &=& -\frac{1}{\xi _y}\int d^5x\Big\vert\partial_{\mu}W^{\mu +}\nonumber \\
&+&\xi_{y}(\partial_{y}W^{4+}+iM_{W}\phi^{+}\{1 + r_{V}\left( \delta(y) + \delta(y - \pi R)\right)\})\Big \vert ^2 \nonumber \\&&-\frac{1}{2\xi_y}\int d^5x [\partial_{\mu}Z^{\mu}\nonumber \\
&+&\xi_y(\partial_{y}Z^{4}-M_{Z}\chi\{1+ r_V( \delta(y) +  \delta(y - \pi R))\})]^2\nonumber \\
&-&\frac{1}{2\xi_y}\int d^5x [\partial_{\mu}A^{\mu}+\xi_y\partial_{y}A^{4}]^2. 
\label{gauge-fix}
\end{eqnarray}\\


The above gauge fixing action is somewhat special and at the same time very crucial for this NMUED scenario. The presence of the BLKTs in the Lagrangian lead to a non-homogeneous weight function for the fields with respect to the extra dimension. This inhomogeneity forces us to choose a $y$-dependent gauge fixing parameter $\xi_y$ as \cite{zbb, bmm, bsg, Dey:2016cve, gf},

\begin{equation}
\xi =\xi_y\,(1+ r_V\{ \delta(y) +  \delta(y - \pi R)\}),
\end{equation}
here $\xi$ is independent of $y$. This relation can be viewed as {\em renormalisation} of the gauge fixing parameter as the 
BLKTs are in some sense counter terms taking into account the unknown ultraviolet contribution in loop calculations. In this sense, 
$\xi_y$ is the bare gauge-fixing parameter while $\xi$ can be viewed as the renormalized gauge-fixing parameter taking the values $0$ (Landau gauge),
$1$ (Feynman gauge) or $\infty$ (Unitary gauge) \cite{gf}.

Proper gauge fixing necessitates $r_V=r_\phi$~\cite{zbb, bmm, bsg,  Dey:2016cve, gf}. As a consequence KK-masses for the scalar and gauge field are equal ($m_{\phi^{(n)}}(=m_{V^{(n)}})$) and follow the same transcendental equation (eq.~\ref{fermion_mass}). Mass eigenvalue for the $n^{th}$ KK-mode of gauge fields ($W^{\mu (n)\pm}$) and charged Higgs ($H^{(n)\pm}$) is \cite{zbb, bmm, bsg, Dey:2016cve, gf} 
\be\label{MWn}
M_{W^{(n)}} = \sqrt{M_{W}^{2}+m^2_{V^{(n)}}}.
\ee

To this end we would like to discuss the necessary interactions that are relevant for our calculation. In general we obtain these by integrating out the 5D action over the extra space-like dimension  after substituting the $y$-dependent KK-wave function for the respective fields in the 5D action. Consequently some of the MUED counter parts are scaled by the so called overlap integrals~\cite{bmm,bsg}. Furthermore in NMUED model we have several extra interacting vertices (which contain the overlap integral) with respect to the MUED model. We provide the overlap integrals crucial for our analysis below:

(i) The interaction between a pair of zero-mode (left-handed) fermion and $n^{th}$ ($n$ being the non-zero even KK-number) KK-mode of $W$-boson:
\begin{eqnarray}
\label{gauge_overlap1}
I^g_n&=&\sqrt{\pi R\left(1+\frac{r_V}{\pi R}\right)}\times \nonumber \\
                &&\displaystyle \int_0 ^{\pi R}\hspace*{-.5cm}dy\left[1 + r_{f}\{ \delta(y) + \delta(y - \pi R)\}\right]a^nf_L^0f_L^0,
\label{gff}
\end{eqnarray}
where $a^n$ is the wave function for $n^{th}$ KK-mode of the $W$-boson.

(ii) Yukawa interaction between a pair of zero-mode fermion and $n^{th}$ ($n$ being the non-zero even KK-number) KK-mode of scalar:
\begin{eqnarray}
\label{youkawa_overlap}
I^Y_n&=&\sqrt{\pi R\left(1+\frac{r_V}{\pi R}\right)}\times \nonumber \\ 
&&\displaystyle \int_0 ^{\pi R}\hspace*{-.5cm}
dy\left[1 + r_{y}\{ \delta(y) + \delta(y - \pi R)\}\right]h^nf_L^0g_R^0.
\end{eqnarray}
In eq.~\ref{youkawa_overlap}, $h^n$ is the wave function for the $n^{th}$ KK-mode of the scalar field. In our case we set $r_{\phi}=r_V$ hence $h^n\equiv a^n$  and further from the equality condition ($r_f=r_y$) one obtains $I^g_n\equiv I^Y_n$ \cite{bmm, bsg}. Without any loss of generality we call it $I_n$. The corresponding expression is:
\beq
I_n= \frac{\sqrt{2}\sqrt{{1+\frac{r_V}{\pi R}}}}{\left({1+\frac{r_f}{\pi R}}\right){\sqrt{1 + \frac{r^2_V m^2_{V^{(n)}}}{4} + \frac{r_V}{\pi R}}}}\frac{\left(r_{f} - r_{V}\right)}{\pi R}.
\label{i1}
\eeq

This integral becomes zero in MUED model due to the orthogonality property of the wave functions of the KK-fields. Hence the $b\rightarrow c \ell \nu_{\ell}$ transitions relevant for our present article that are mediated by KK-$W$-bosons and KK-charged scalar fields will not occur in the MUED model. However for the NMUED model, the above overlap integral appears in the vertices involving the $b\rightarrow c \ell \nu_{\ell}$ decay amplitudes. One should note that $I_n$ vanishes for $r_V=r_f$.

\section{$\mathcal{R}(D)$ and $\mathcal{R}(D^*)$}
\begin{table}[!hbt]
\begin{center}
\renewcommand{\arraystretch}{1.2}
\begin{tabular}{ccccccc}
\hline
\hline
Parameters 	& Value		&\multicolumn{5}{c}{Correlation}\\
\hline
 $\rho_D^2$ 	& 1.075(42) 	& 1. 	& 0.26 		& -0.01 	& -0.13 	& 0 \\
 $\rho_{D^*}^2$ & 1.221(118)	&  	& 1. 		& 0.08 		& -0.80 	& 0 \\
 $R_1(1)$ 	& 1.372(36) 	&  	& 		& 1. 		& -0.08 	& 0.21 \\
 $R_2(1)$ 	& 0.895(65) 	&  	& 		& 		& 1. 		& -0.01 \\
 $R_0(1)$ 	& 1.186(16) 	&  	&  		&  		&  		& 1 \\
 \hline
 $m_B$		& 5.27962(15) GeV	& \\
 $m_{D^*}$	& 2.01026(5) GeV	& \\
 $m_W$ 		& 80.385(15) GeV	& \\
 $m_W$ 		& 80.385(15) GeV	& \\
 $m_c$ 		& 1.28(3) GeV	& \\
 $m_b$ 		& $4.18^{+0.04}_{-0.03}$ GeV	& \\
 $m_{\tau}$	& 1.77682(16) GeV	& \\
\hline
\hline
\end{tabular}
\caption{Nuisance inputs in the theory expressions. Only those form factor parameters which appear in $\mathcal{R}(D^{(*)})$ are shown here (with correlations). These are obtained from the analysis in ref. \cite{Jaiswal:2017rve}.} 
\label{tab:nuisance}
\end{center}
\end{table}

\begin{table*}[!hbt]
\begin{center}
\renewcommand{\arraystretch}{1.2}
\begin{tabular}{cccc}
\hline
\hline
			& $\mathcal{R}(D)$                      	& $\mathcal{R}(D^*)$  & Correlation\\
\hline
\noalign{\vskip1pt}
SM  		   	& $0.300(8)$ \cite{Na:2015kha}   	& $0.252(3)$ \cite{Fajfer:2012vx} 		&\\
			& $0.299(11)$ \cite{Lattice:2015rga}	& 						&\\
			& $0.299(3)$ \cite{Bigi:2016mdz}	& $0.262(10)$ \cite{Bigi:2017jbd}		&\\
			& $0.299(3)$				& $0.257(3)$ 		& $0.44$ \cite{Bernlochner:2017jka}\\
			& ${\bf 0.302(3)}$			& ${\bf 0.257(5)}$ 	& ${\bf 0.127}$ \cite{Jaiswal:2017rve}\\
			\hline
\Babar~   	 	& $0.440(58)_{st.}(42)_{sy.} $ 	 	& $0.332(24)_{st.}(18)_{sy.}$ & $-0.27$\cite{Lees:2013uzd}\\
Belle (2015)  	 	& $0.375(64)_{st.}(26)_{sy.}$ 	 	& $0.293(38)_{st.}(15)_{sy.}$ & $-0.49$\cite{Huschle:2015rga}\\
Belle (2016)  	 	& -			 		& $0.302(30)_{st.}(11)_{sy.}$ \cite{Abdesselam:2016cgx} & \\
Belle (2016,		& - 					& $0.270(35)_{st.}~^{+ 0.028}_{-0.025} $ \cite{Hirose:2016wfn} & \\
Full Dataset) 		& & & \\
LHCb (2015)    		& - 					& $0.336(27)_{st.}(30)_{sy.}$ \cite{Aaij:2015yra} & \\
LHCb (2017)    		& - 					& $0.285(19)_{st.}(29)_{sy.}$ &(Presented at FPCP2017)\\
\hline
World Avg.		& $0.407(39)_{st.}(24)_{sy.}$		& $0.304(13)_{st.}(7)_{sy.}$ 	& $0.20$ \cite{hfag_rdrdst}\\
\hline
\hline
\end{tabular}
\caption{Present status (both theoretical and experimental) of $\mathcal{R}(D)$ and $\mathcal{R}(D^*)$. The first uncertainty is the statistical one and the second one is systematic. The SM calculation in this paper is closest to the one in bold-face.} 
\label{tab:RDRDst}
\end{center}
\end{table*}
%

\subsection{Formalism}\label{rdrdst_formalism}

\begin{figure*}\centering
\subfloat[All Data]{
\includegraphics[scale=0.4]{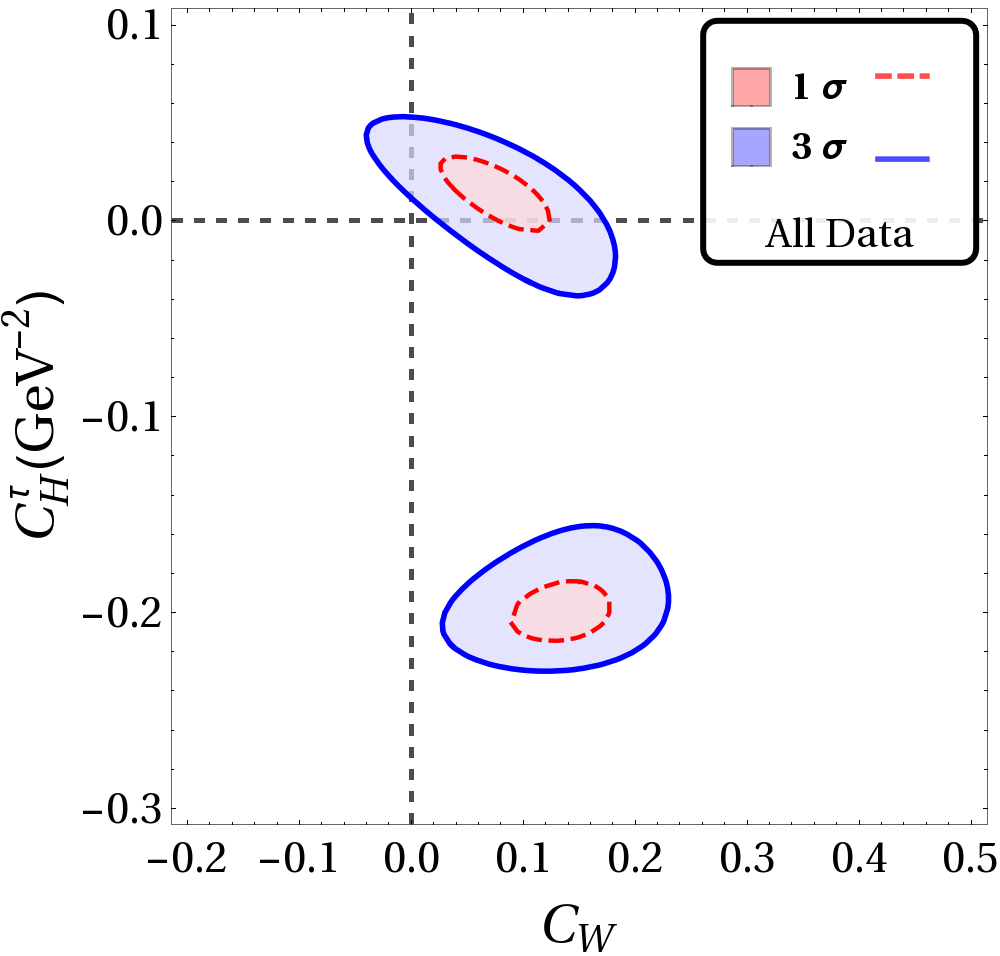}
 \label{fig:plt1}}
\subfloat[All Belle]{
\includegraphics[scale=0.4]{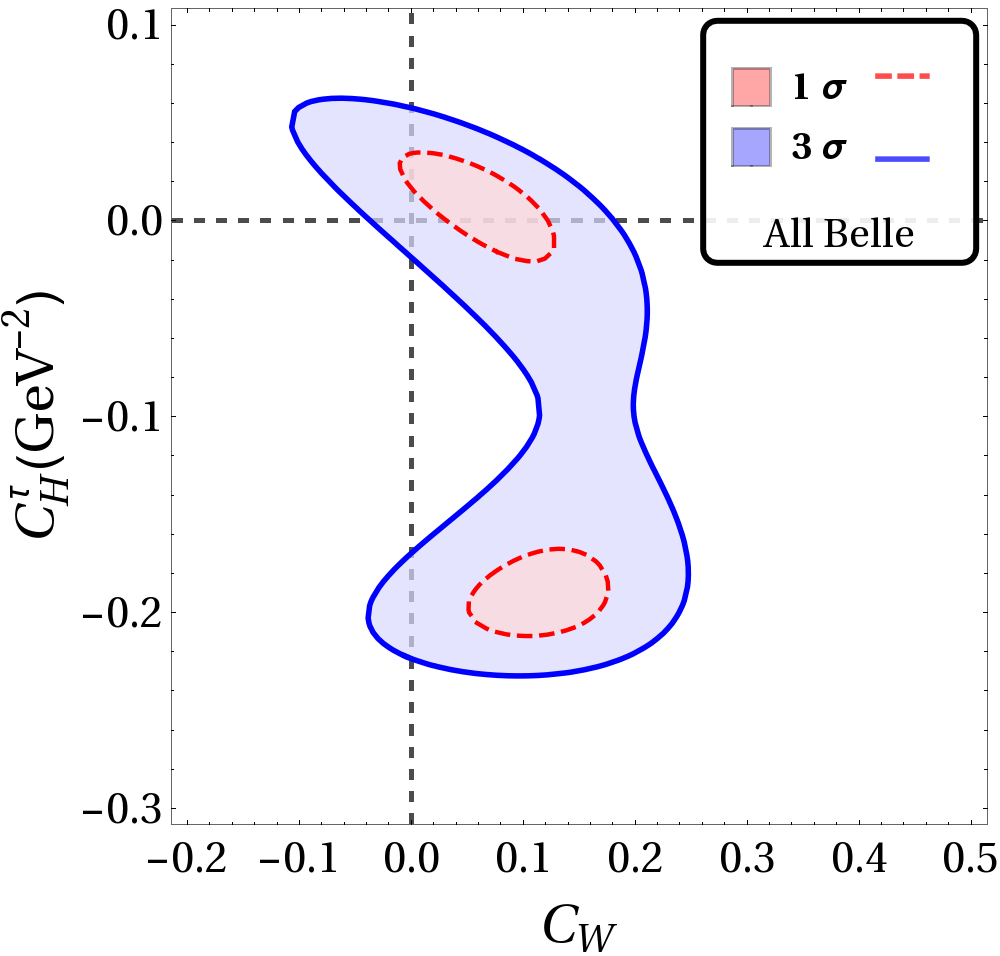}
 \label{fig:plt2}}
\subfloat[\Babar~+ LHCb]{
\includegraphics[scale=0.4]{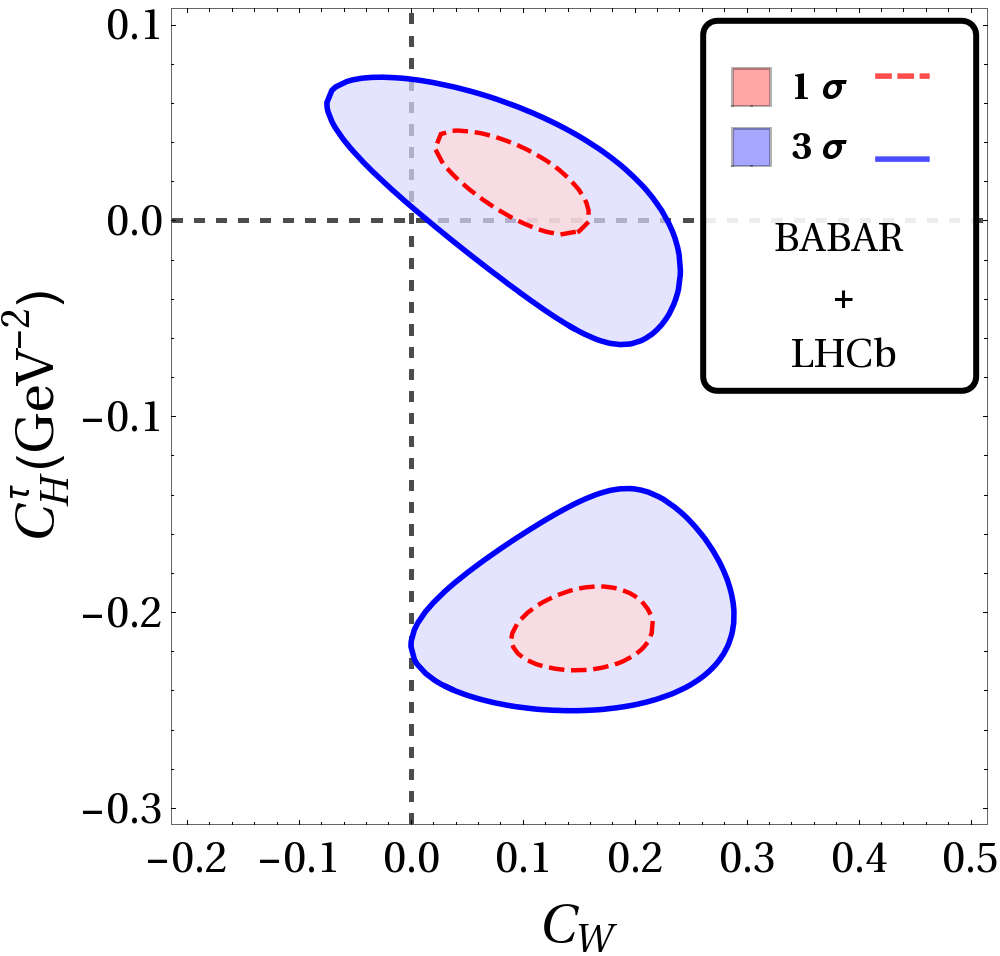}
 \label{fig:plt3}}\\
\subfloat[\Babar~+ Belle]{
\includegraphics[scale=0.4]{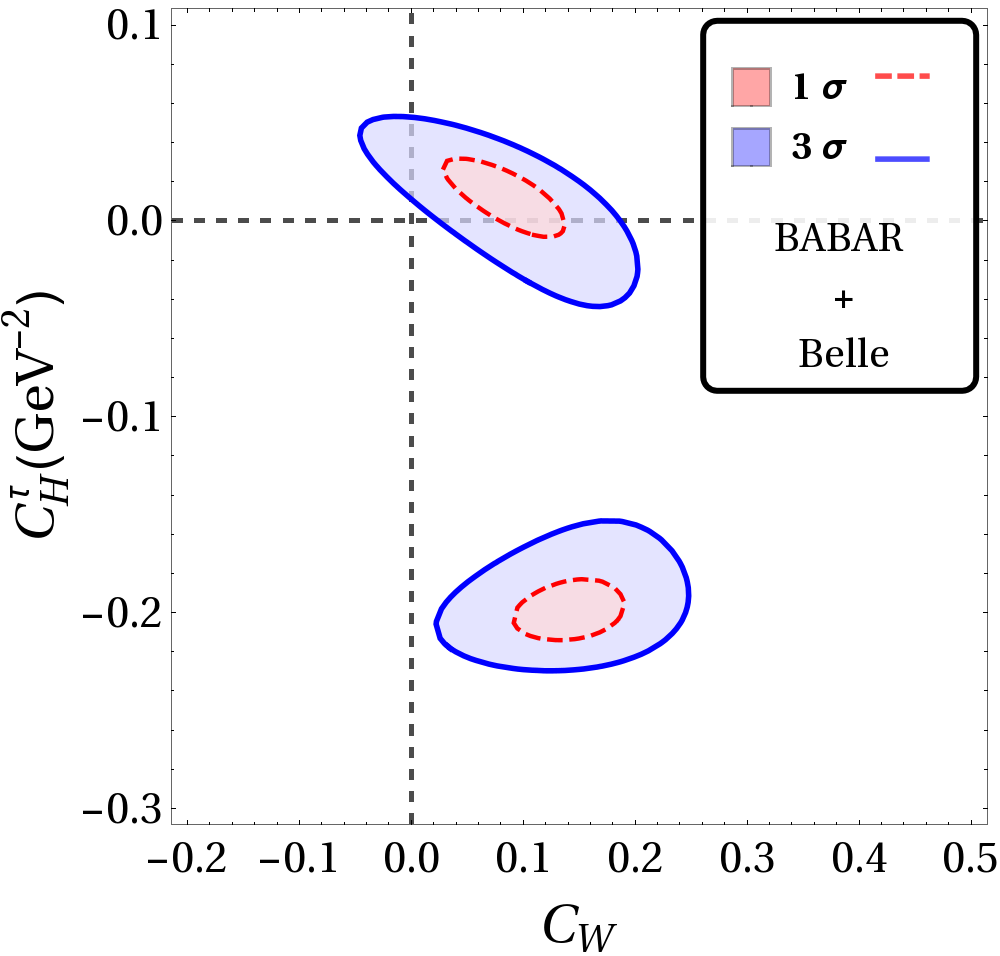}
 \label{fig:plt4}}
\subfloat[Belle + LHCb]{
\includegraphics[scale=0.4]{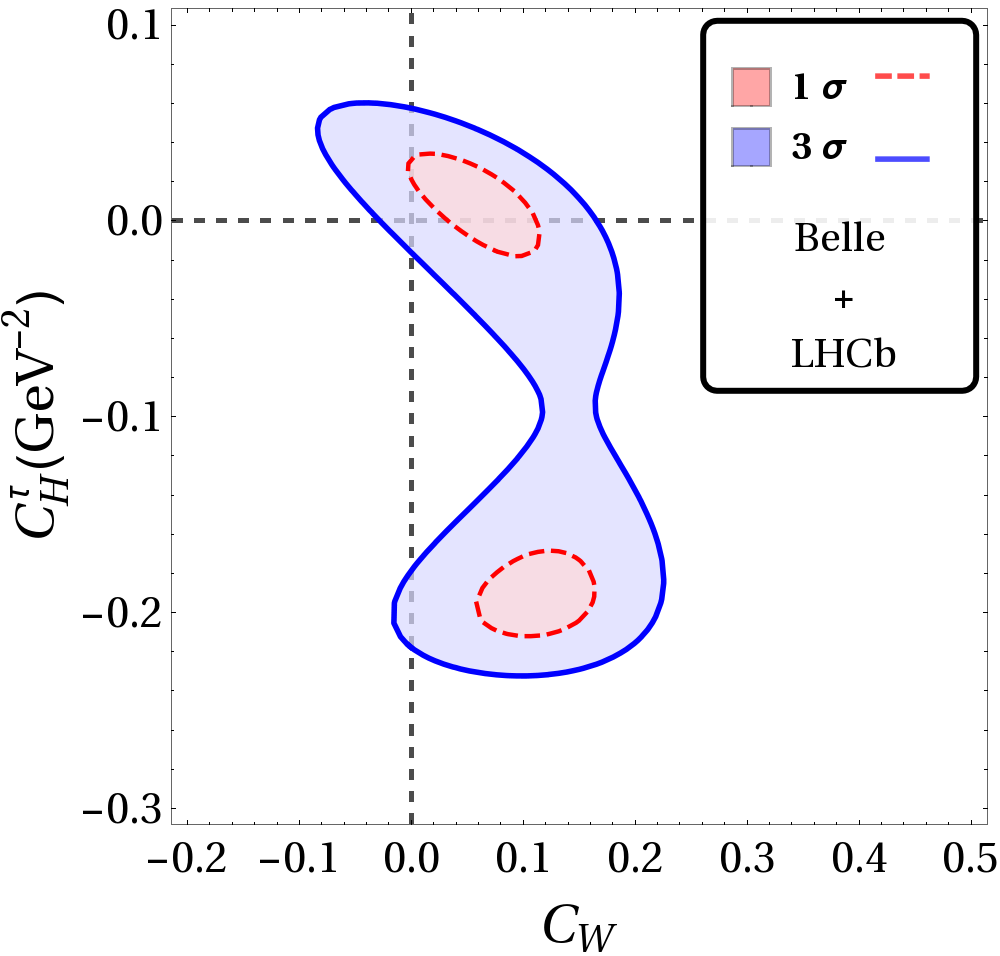}
 \label{fig:plt5}}
\subfloat[All Except Latest $\mathcal{R}(D^{*})$]{
\includegraphics[scale=0.4]{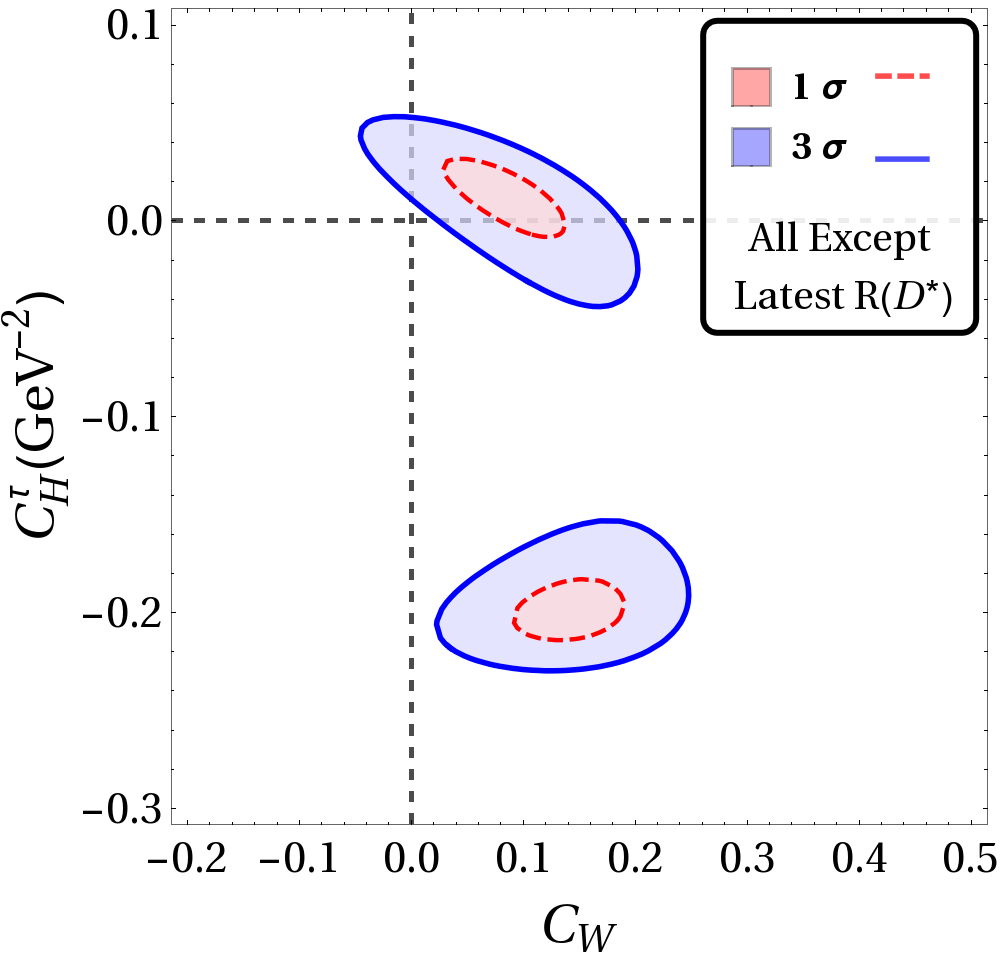}
 \label{fig:plt7}}
 \caption{$\mathcal{R}(D^{(*)})$ fit results corresponding to separate fits listed in table \ref{tab:res1}. Red(dotted) and blue(solid) lines enclose $1 \sigma$ ($\Delta\chi^2 = 2.30$) and $3 \sigma$ ($\Delta\chi^2 = 11.83$) regions respectively. Only the gridlines corresponding to $C_W$ and $C^{\tau}_H = 0$ are shown, such that there intersection point represents SM.}
\label{fig:chiplts}
\end{figure*}

In presence of NMUED, the general effective Hamiltonian describing the $b\to c\tau \nu_{\tau}$ transitions with 
all possible four-fermi operators in the lowest dimension is given by 
\begin{equation}
{\cal H}_{eff} = \frac{4 G_F}{\sqrt{2}} V_{cb}\Big[( 1 + C_{W}) {\cal O}_{W} + C^{\ell}_{S} {\cal O}_{S}\Big],
 \label{effH}
\end{equation}
where, following the convention of \cite{Sakaki:2014}, the operator basis is defined as
\begin{eqnarray}
{\cal O}_{W} &=& ({\bar c}_L \gamma^\mu b_L)({\bar \tau}_L \gamma_\mu \nu_{\tau L}) \nn,\\
{\cal O}_{S} &=& ({\bar c}_L  b_R)({\bar \tau}_R \nu_{\tau L}) \nn \,,
\label{ops}
\end{eqnarray}
and the corresponding Wilson coefficients are given by $C_X (X=W,~S)$. Following ref. \cite{Bhattacharya:2016zcw} and references therein, differential decay rates for $\bar{B}\rightarrow D^{(*)} \ell \bar{\nu_\ell}$ (with $\ell = e$, $\mu$ or $\tau$ ) are 
\vspace*{1cm}
\begin{widetext}
\begin{align}
\nn \frac{d\Gamma(\bar{B}\rightarrow D \ell \bar{\nu_\ell})}{dq^2} &= \frac{G^2_F |V_{cb}|^2}{96 \pi^3 m^2_B}q^2 p_D \left(1- \frac{m^2_\ell}{q^2}\right)^2\bigg[(1+C_W)^2  \left(1+ \frac{m^2_\ell}{2q^2}\right)^2 H^{s 2}_{V, 0}  \\
 &+ \frac{3 m^2_\ell}{2 q^2} H^{s 2}_{V, t}(1 + C_W + \frac{q^2}{m_{\ell}\left(m_b - m_c\right)} C^{\ell}_{S})^2 \bigg]\\
\nn \frac{d \Gamma(\bar{B}\rightarrow D^* \ell \bar{\nu_\ell})}{dq^2} &= \frac{G^2_F|V_{cb}|^2}{96(\pi)^3 m^2_B}q^2p_{D^*}\left(1- \frac{m^2_\ell}{q^2}\right)^2 \left[(1+C_W)^2(1+\frac{m^2_{\ell}}{2 q^2})\left(H^2_{V, +} + H^2_{V, -} + H^2_{V, 0}\right) \right.\\
 &\left.+ \frac{3 m^2_{\ell}}{2 q^2} (1 + C_W + \frac{q^2}{m_{\ell}\left(m_b + m_c\right)} C^{\ell}_{S})^2 H^2_{V, t}\right]\,,
\label{rdrds}
\end{align}
\end{widetext}
%

where
$p_{D(D^*)}=\frac{\lambda^{1/2} (m^2_B,~m^2_{D(D^*)},~q^2)}{2m_B}$, with $\lambda(a~,b,~c)=a^2+b^2+c^2-2(ab+bc+ca)$ and $q_\mu \equiv (p_B - p_X)_\mu$ is the momentum transfer. $H^s_{V, X}(q^2)$ and $H_{V, X}(q^2)$ are the helicity amplitudes respectively (with $X = \pm, ~0$).

In terms of these distributions, the ratios $\mathcal{R}(D)$ and $\mathcal{R}(D^*)$ are defined as
\begin{align}\label{Rth}
\nn \mathcal{R}(D^{(*)}) &= \left[\int^{q^2_{max}}_{m^2_{\tau}} \frac{d\Gamma\left(\overline{B} \rightarrow D^{(*)} 
 \tau \overline{\nu}_{\tau}\right)}{d q^2} d q^2\right]\times \\
&\left[\int^{q^2_{max}}_{m^2_{\ell}} 
 \frac{d\Gamma\left(\overline{B} \rightarrow D^{(*)} \ell \overline{\nu}_{\ell}\right)}{d q^2} d q^2\right]^{-1}
\end{align}
with $q^2_{max}= (m_B - m_{D^{(*)}})^2$, and $\ell=e$ or $\mu$. In both cases, both isospin channels are taken into account.

Let us spend some time on the  Wilson coefficients and their expressions which are relevant in our present article. First, we discuss the Wilson coefficient $C_W$ (given in eq.\;\ref{effH}) which is associated with the extra left handed charged currents in the gauge sector. The expression is given in the following:
\beq
C_W=\sum_{n \geq 2}{\frac{I^2_n M^2_W}{M^2_{W^{(n)}}}}~.
\label{cw}
\eeq
This originates from the coupling between pair of SM fermions (quark or lepton) with the KK excited $W^\pm$ boson (i.e., $W^{\pm(n)} q \bar{q'}$ and $W^{\pm(n)} \ell \nu_{\ell}$) (see eq.\;\ref{factn}). As we have chosen the same BLKT coefficient ($r_f$) for all fermions, so the coupling of $W^{\pm(n)}$ to $\ell \nu_{\ell}$ is same for all lepton flavour \cite{ddrs2, lhc, zbb, bmm, bsg, Dey:2016cve, ddrs1}. Therefore, there is no lepton flavor
universality (LFU) violation in the gauge sector. However, LFU violation is possible via another Wilson coefficient $C^{\ell}_S$ (associated with left-handed scalar
type NP charged current interactions) whose expression is given below
\begin{align}
\nn C^{\ell}_S &= m_b m_{\ell}\sum_{n \geq 2}{\frac{I^2_n m^2_{V^{(n)}}}{M^4_{W^{(n)}}}}\times\\
\nn &[\cos(c(n)-\ell(n)) -\sin(c(n)+\ell(n))]\\
&\equiv m_b m_{\ell} C^{\ell}_{H}\,, ~~~~~(\ell \equiv e,~\mu~ {\rm or } ~\tau).
\label{chl}
\end{align}
This is generated from the interaction given in eq.\;\ref{yukawa}. The explicit form of the couplings of this interaction are given in the appendices of refs.\;\cite{bmm, bsg}. We find that the coupling  $H^{(n)\pm}$ to $\ell \nu_{\ell}$ is lepton flavour dependent by means of lepton mass and using those couplings we calculate the Wilson coefficient $C^{\ell}_S$ (see eq.\;\ref{chl}). Hence from this $C^{\ell}_S$ we obtain the LFU violation which is very crucial for the present $\mathcal{R}(D^{(*)}$ analysis.

The expressions of $M_{W^{(n)}}$ and $I_n$ are given by eqs.~\ref{MWn} and~\ref{i1} respectively. Here, $n$ is a non-zero even integer. Using eq.~\ref{alphatn} one can obtain $c(n)$ and $\ell(n)$ from the following equations:
\begin{align}
\tan[2 c(n)] &= \frac{m_c}{m_{f^{(n)}}},\\
\tan[2 \ell(n)] &= \frac{m_{\ell}}{m_{f^{(n)}}},
\end{align}
where, $m_c$ is the mass of charm quark and $m_{\ell}$ denotes the mass of charged lepton.

\begin{figure}
 \includegraphics[scale=0.4]{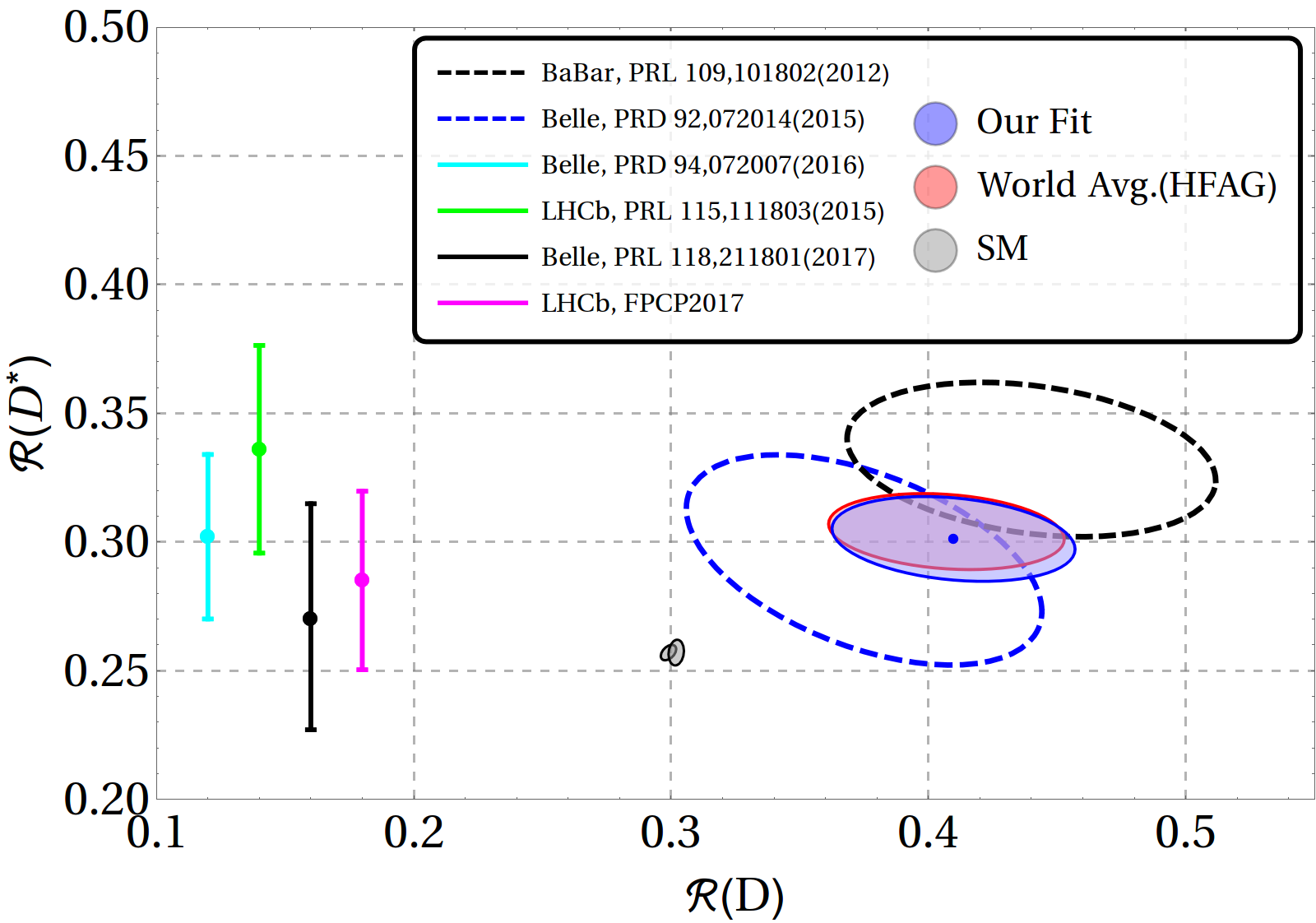}
 \caption{Experimental values and NMUED prediction (with $1\sigma$ errors) for the $\mathcal{R}(D)$ and $\mathcal{R}(D^*)$ ratios represented on the $\mathcal{R}(D)$ vs $\mathcal{R}(D^*)$ plane.}
 \label{rdrdstplt}
\end{figure}

\subsection{Fit of $C_W$ and $C^{\tau}_H$}\label{chwfit}

\subsubsection{Present Status}

Till date, several experiments have measured the ratios $\mathcal{R}(D^{(*)})$, and the current status is summarized in Table \ref{tab:RDRDst} and fig.~\ref{rdrdstplt} where the results show the $q^2$-integrated data on $\mathcal{R}(D)$ and $\mathcal{R}(D^*)$ with appropriate
correlations wherever the data is available. Though both the \Babar~and the Belle collaborations have also
published the results on differential distributions, which would increase the sensitivity only nominally, we refrain from using the binned 
data. The reasons are: 
\begin{enumerate}[label=\alph*)]
 \item While all data, apart from Belle 2015 and the latest LHCb results, are consistent with a sizable deviation
from the SM expectations, there is some tension between the $q^2$ distributions as seen by \Babar~ and Belle. As a result, using this data would not lead to any significant improvement in the results.
  \item While the binned data by \Babar~is independent of the background models, Belle 2015 data is noticeably not. The SM expectations used by the two collaborations also differ from each other.
  \item No other result accompanies the $q^2$ binned data.
\end{enumerate}

\subsubsection{Methodology}

As the NMUED model parameters $R_V$, $R_f$ and $R^{-1}$ occur in transcendental equations like eq.~\ref{fermion_mass}, fitting them directly from the experimental data of $\mathcal{R}(D^{(*)})$ is deemed improbable. We instead fit $C_W$ and $C^{\tau}_H$ and then find out the allowed parameter space for the model parameters from the fit results. We constrain both parameters to be real and $C_W$ to be positive. 

For the theoretical expressions of $\mathcal{R}(D^{(*)})$, we follow the Caprini-Lellouch-Neubert (CLN) \cite{Caprini:1997mu} parametrization of the $B\to D^{(*)}$ form factors. Table \ref{tab:nuisance} contains the full information on the nuisance parameters. As $m_e$ or $m_{\mu}$ are very small compared to $m_{\tau}$, their effect in $\mathcal{R}(D^{(*)})$ would be negligible, at least in the present work.
To fit $C_W$ and $C^{\tau}_H$, we have performed a test of significance (goodness of fit) by defining a $\chi^2$ statistic, a function of the parameters, which is defined as
\begin{align}\label{chidef}
 \nn \chi^2_{\rm NMUED} &= \sum^{{\rm data}}_{i,j = 1} \left(\mathcal{R}(D^{(*)})^{exp}_i - \mathcal{R}(D^{(*)})^{th}\right)\\
 \nn &\left(V^{stat} + V^{syst} \right)^{-1}_{i j} \left(\mathcal{R}(D^{(*)})^{exp}_j - \mathcal{R}(D^{(*)})^{th}\right)\\
 &+ \chi^2_{Nuisance}\,,
\end{align}
where $\mathcal{R}(D^{(*)})^{th}$ is given by eq.~\ref{Rth} and $\mathcal{R}(D^{(*)})^{exp}_i$ is the central value of the $i$th experimental result. Also,
\begin{align}
\nn \chi^2_{Nuisance} &= \sum^{{\rm theory}}_{i,j = 1} \left(param_i - value_i\right) \left(V^{nuis}\right)^{-1}_{i j}\\ 
 &\left(param_j - value_j\right)\,.
\end{align}
In all of these cases, $V_{i j} = \delta_i\times \delta_j \times \rho_{i j}$ is the respective covariance matrix, where $\rho_{i j}$ is the correlation between $i$th and $j$th observable or parameter.

\begin{figure}
 \includegraphics[scale=0.5]{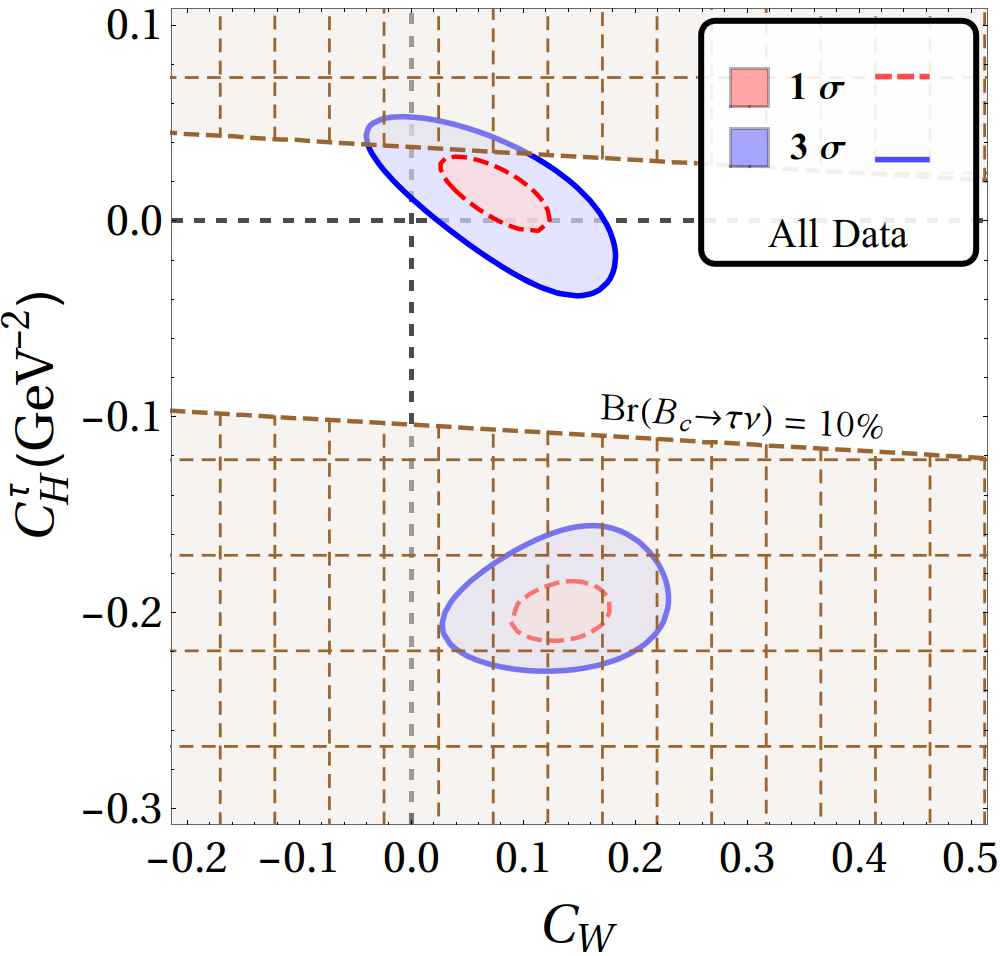}
 \caption{Fig.~\ref{fig:plt1} overlaid with the region excluded by demanding ${\rm Br}(B^-_c \to \tau^- \bar{\nu}) \lesssim 10\%$.}
 \label{pltbc}
\end{figure}

\subsubsection{Results}

\begin{table*}[!hbt]
\begin{center}
\renewcommand{\arraystretch}{1.2}
\begin{tabular}{ccccccc}
\hline
\hline
 Data			&$\chi^2_{min}$	& d.o.f	& $p$-value	& $C_W$		& $C^{\tau}_H$ (in GeV$^{-2}$)	& Correlation\\
\hline
 \textbf{All Data}	& \textbf{2.935}& \textbf{6} 	& \textbf{81.694} 	& \textbf{0.076(32)} 	& \textbf{0.015(12)} 	& \textbf{-0.702} \\
 All Belle 		& 0.349 	& 2 	& 83.98 	& 0.060(46) 	& 0.010(18) 	& -0.715 \\
 \Babar~ + LHCb 	& 1.057 	& 2 	& 58.941 	& 0.091(45) 	& 0.022(17) 	& -0.687 \\
 \Babar + Belle 	& 2.652 	& 4 	& 61.77 	& 0.084(36) 	& 0.013(13) 	& -0.728 \\
 Belle + LHCb 		& 0.398 	& 4 	& 98.264 	& 0.057(39) 	& 0.011(17) 	& -0.678 \\
 All Except Latest LHCb & 2.662 	& 5 	& 75.191 	& 0.084(36) 	& 0.013(13) 	& -0.728 \\
\hline
\hline
\end{tabular}
\caption{Fit of $q^2$-integrated results of $\mathcal{R}(D)$ and $\mathcal{R}(D^*)$ with the parameters defined in eqs.~\ref{cw} and \ref{chl}. The uncertainties and correlations are obtained from the hessian and the $p$-value is obtained for the $\chi^2_{min}$ value under a $\chi^2$ distribution of corresponding d.o.f.} 
\label{tab:res1}
\end{center}
\end{table*}

We have taken several combinations of the eight available $\mathcal{R}(D^{(*)})$ data points while fitting. Table \ref{tab:res1} contains the fit results, best fit values of the fit parameters with their uncertainties and correlations. They show that though all of them are good fits, \Babar~ data has a tension with those of Belle and LHCb. Here we note that the latest LHCb data on $\mathcal{R}(D^*)$ is actually obtained by multiplying the particle data group (PDG) average values of appropriate branching fractions with the actually measured ratio $K_{had}(D^*) = Br(B\to D^* \tau \nu) / Br(B\to D^* \pi\pi\pi)$. This is why we have even fitted for a case with this one data-point dropped and we note that inclusion of this point actually gives us a better fit of the data with our new physics coefficients. Each region plot in fig.~\ref{fig:chiplts} contains the  $1\sigma$ and $3\sigma$ contours in the $C_W$ - $C^{\tau}_H$ plane for a different set of experimental results, that are equivalent to $p$-values of $0.3173$ and $0.0027$, corresponding to confidence levels of $68.27\%$ and $99.73\%$, respectively. For our purpose, each confidence interval corresponds to a particular value of $X = \Delta\chi^2$(i.e.\ $\chi^2 - \chi^2_{min}$) for $d.o.f = 2$ (no. of parameters), such that $p(X|{\rm d.o.f})$ is fixed. As an example, $\Delta\chi^2 = 2.30$ and $11.83$ for $1 \sigma$ and $3 \sigma$ regions respectively in 2 dimensions\footnote{Though $\Delta\chi^2 = 1$ gives $1 \sigma$ region for a single PDF and is needed for quoting uncertainties, it encloses a smaller region than the confidence level of $68.27\%$ for any higher dimensional PDF.}.

\begin{figure*}\centering
\subfloat[Inset: zoomed in.]{
\includegraphics[height=5.7cm]{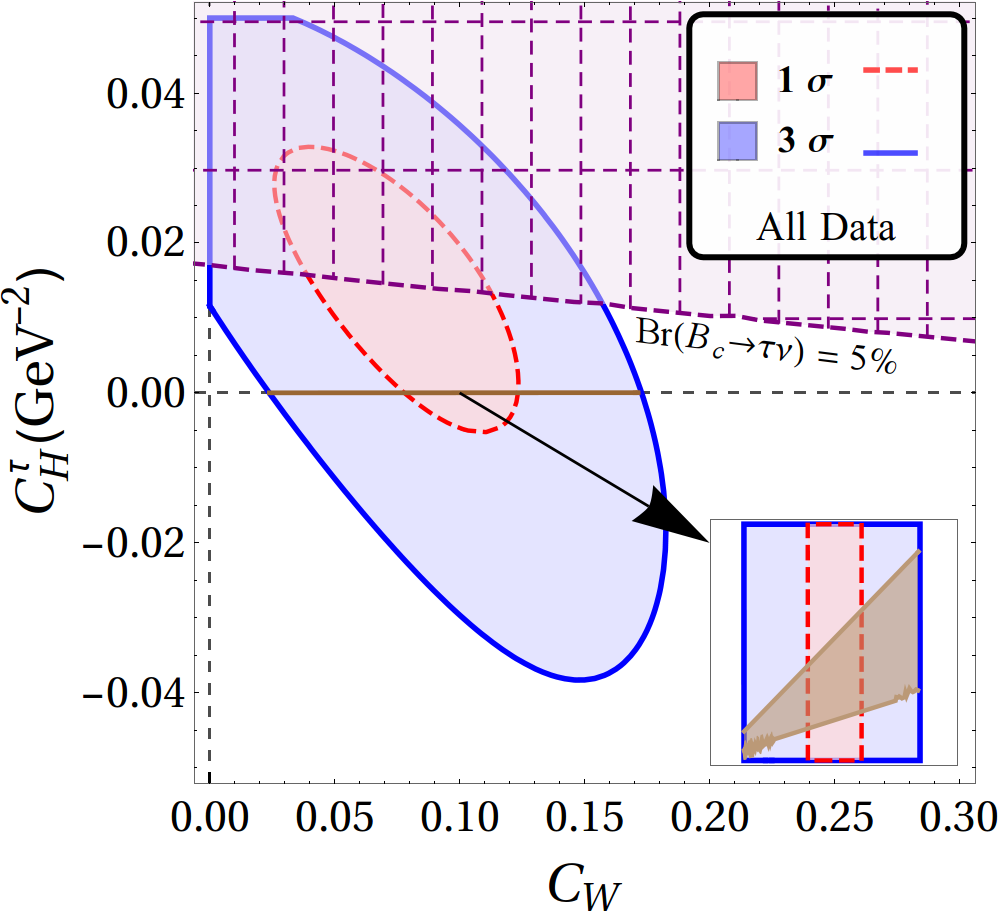}
 \label{fig:pltzoom}}
\subfloat[NMUED allowed space in $C_W$ - $C^{\tau}_H$ plane.]{
\includegraphics[height=5.7cm]{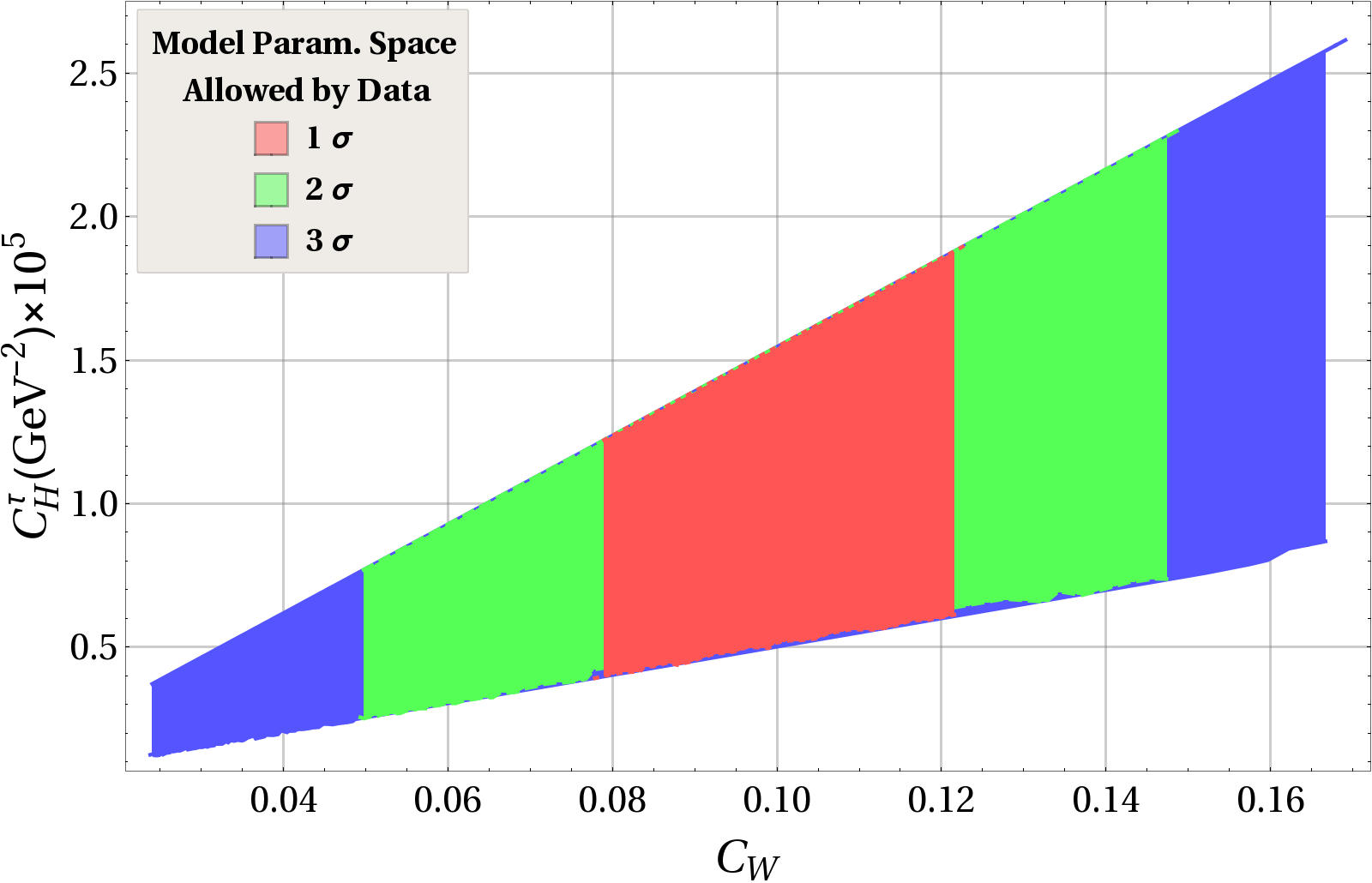}
 \label{fig:cwhpltparam}}
 \caption{Region in $C_W$ - $C^{\tau}_H$ plane, populated by NMUED parameters. In the first figure, the brown (overlay) region points to the allowed parameter space and the purple(hatched) region is excluded by demanding ${\rm Br}(B^-_c \to \tau^- \bar{\nu}) \lesssim 5\%$. The allowed parameter space is blown up in proportion in the second figure.}
\label{fig:modpar1}
\end{figure*}

\begin{figure*}\centering
\subfloat[$R_V$ vs. $R_f$]{
\includegraphics[width=5.5cm, height=5.5cm]{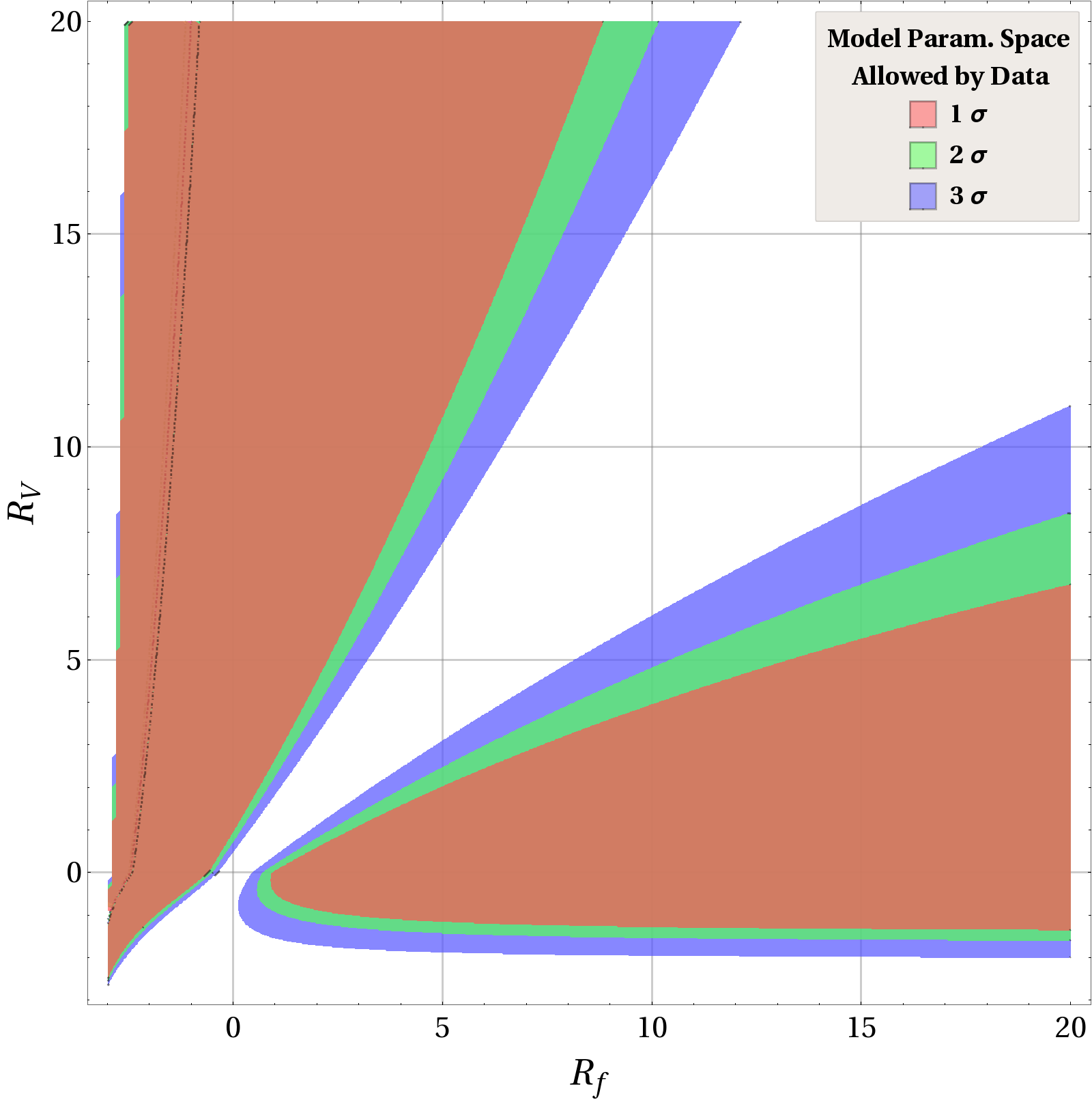}
 \label{fig:rfrvpltparam}}
\subfloat[$R_f$ vs. $R^{-1}$]{
\includegraphics[width=5.5cm, height=5.5cm]{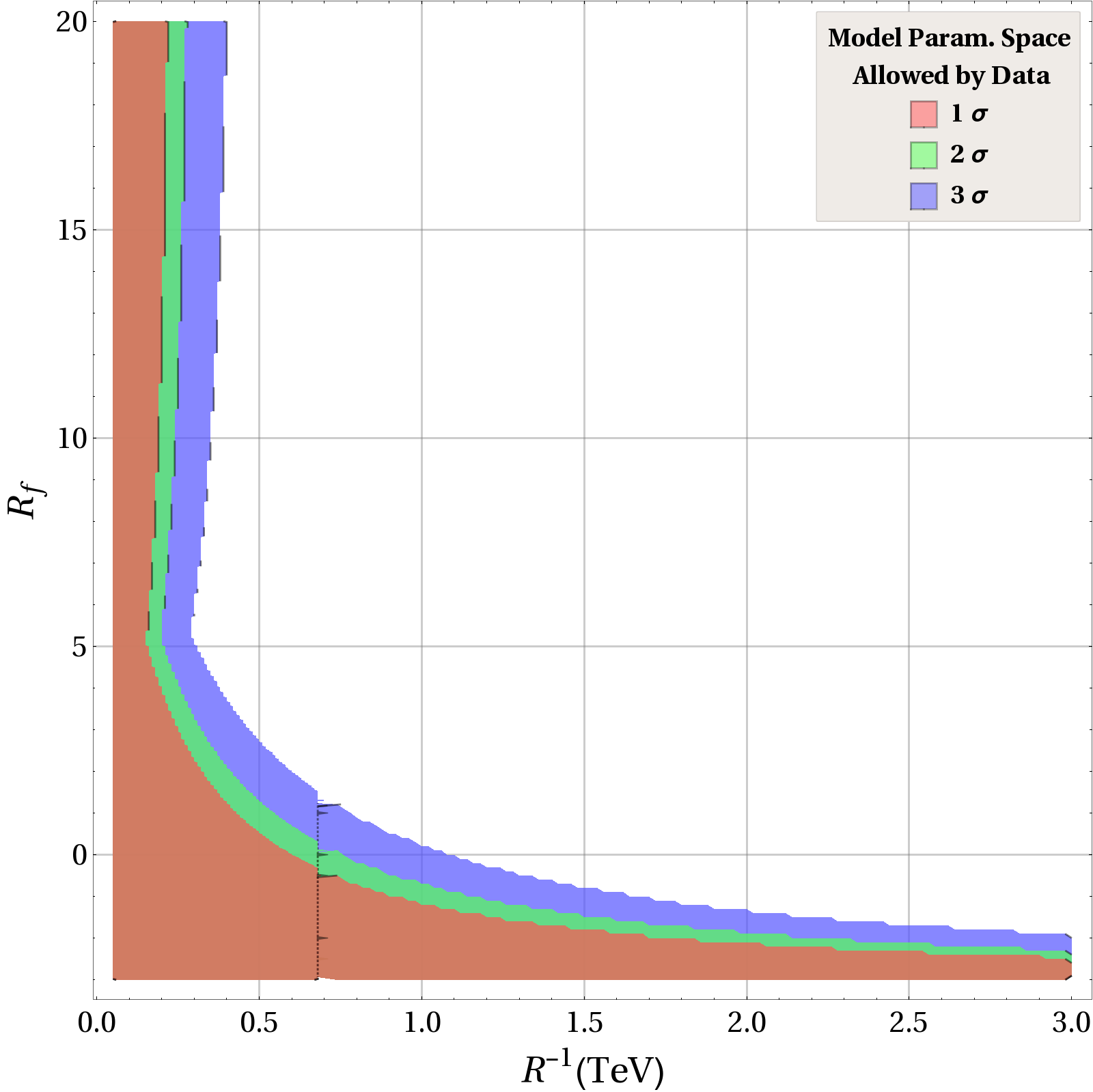}
 \label{fig:rrfpltparam}}
\subfloat[$R_V$ vs. $R^{-1}$]{
\includegraphics[width=5.5cm, height=5.5cm]{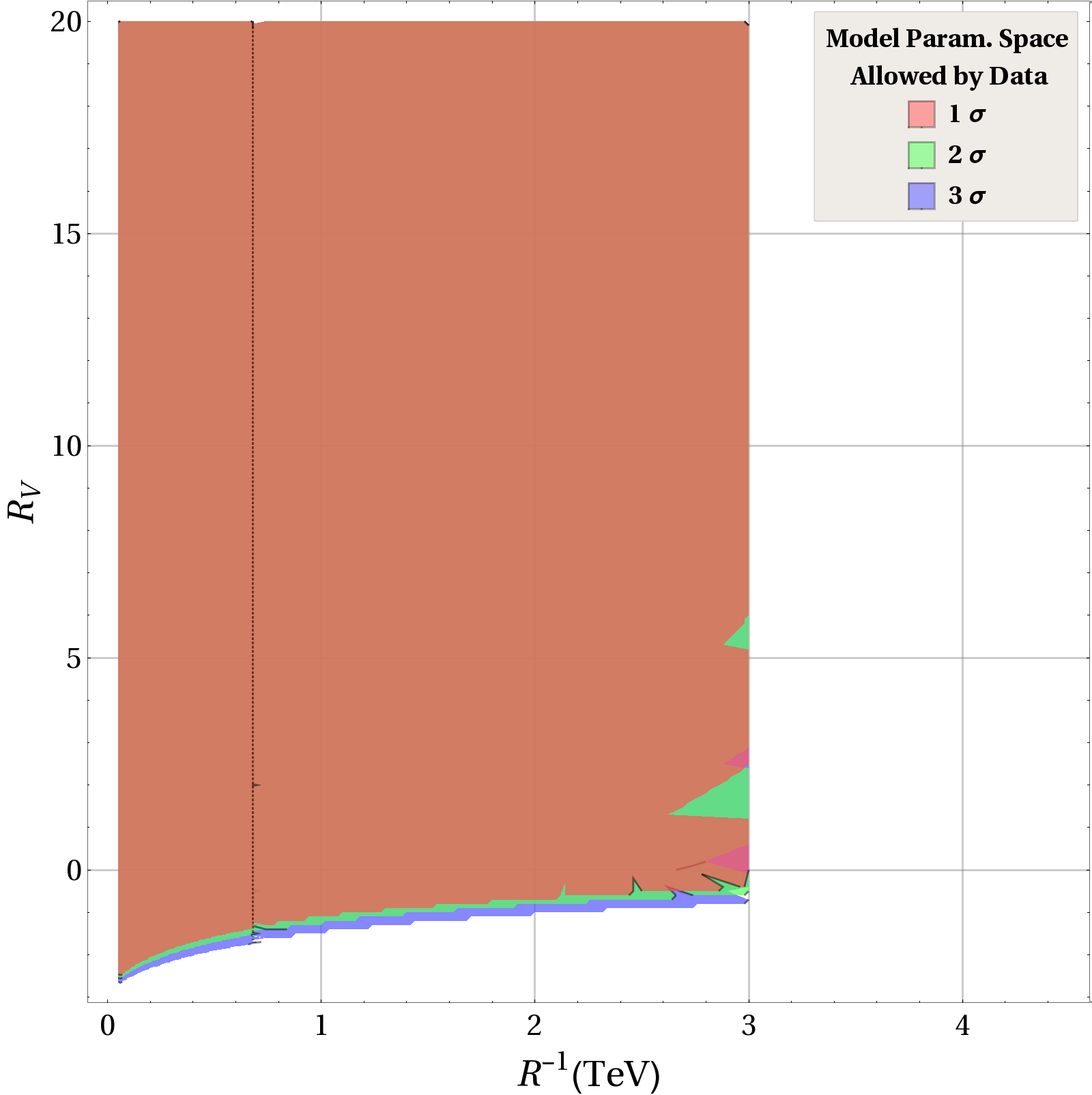}
 \label{fig:rrvpltparam}}
 \caption{Regions in the NMUED model parameter space, allowed by $C_W$ - $C^{\tau}_H$ fit of $\mathcal{R}(D^{(*)})$ data.}
\label{fig:modpar2}
\end{figure*}

\begin{figure*}\centering
\subfloat[$R_V$ vs. $R_f$]{
\includegraphics[width=5.5cm, height=5.5cm]{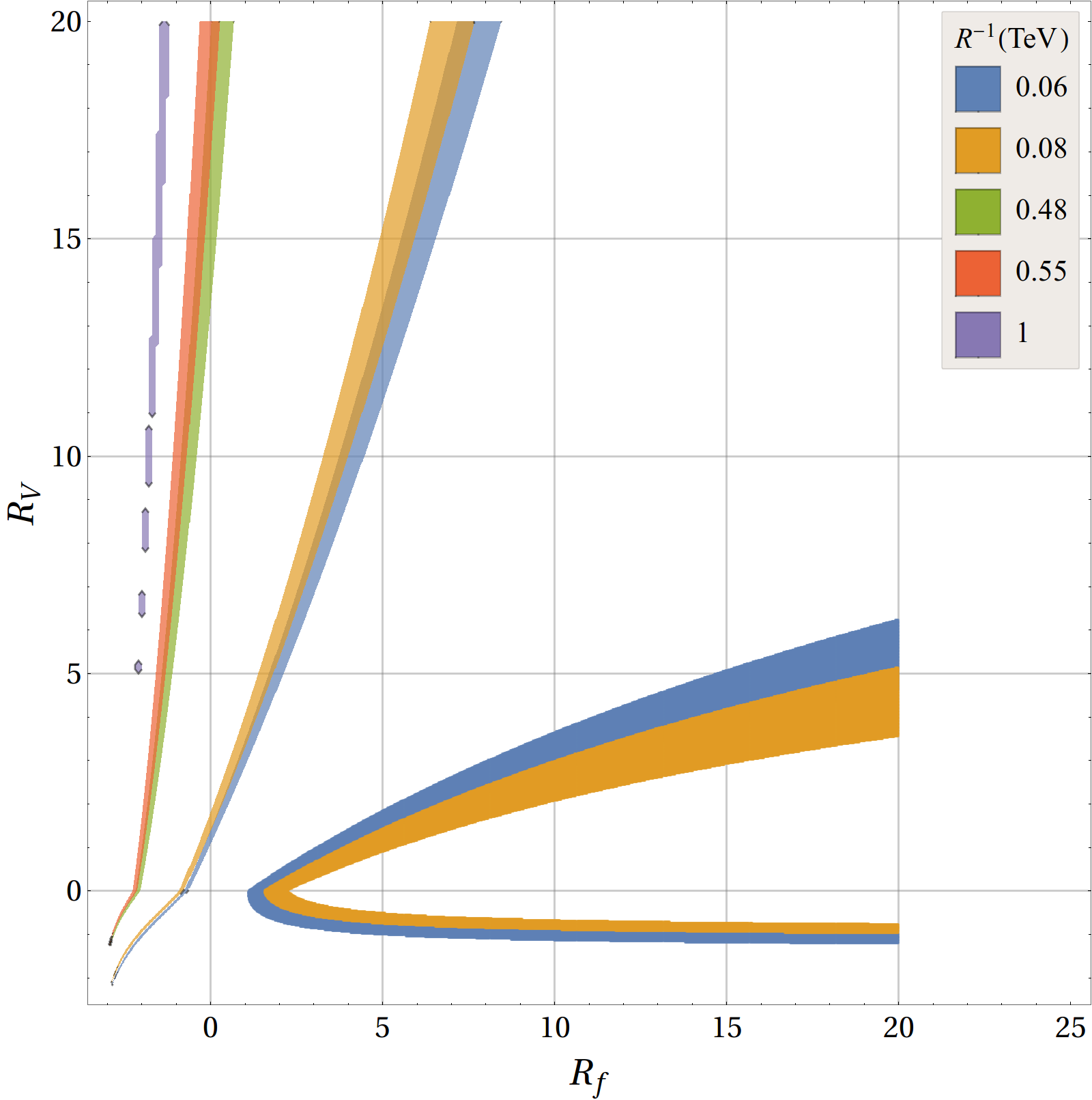}
 \label{fig:rfrvdatplt}}
\subfloat[$R_f$ vs. $R^{-1}$]{
\includegraphics[width=5.5cm, height=5.5cm]{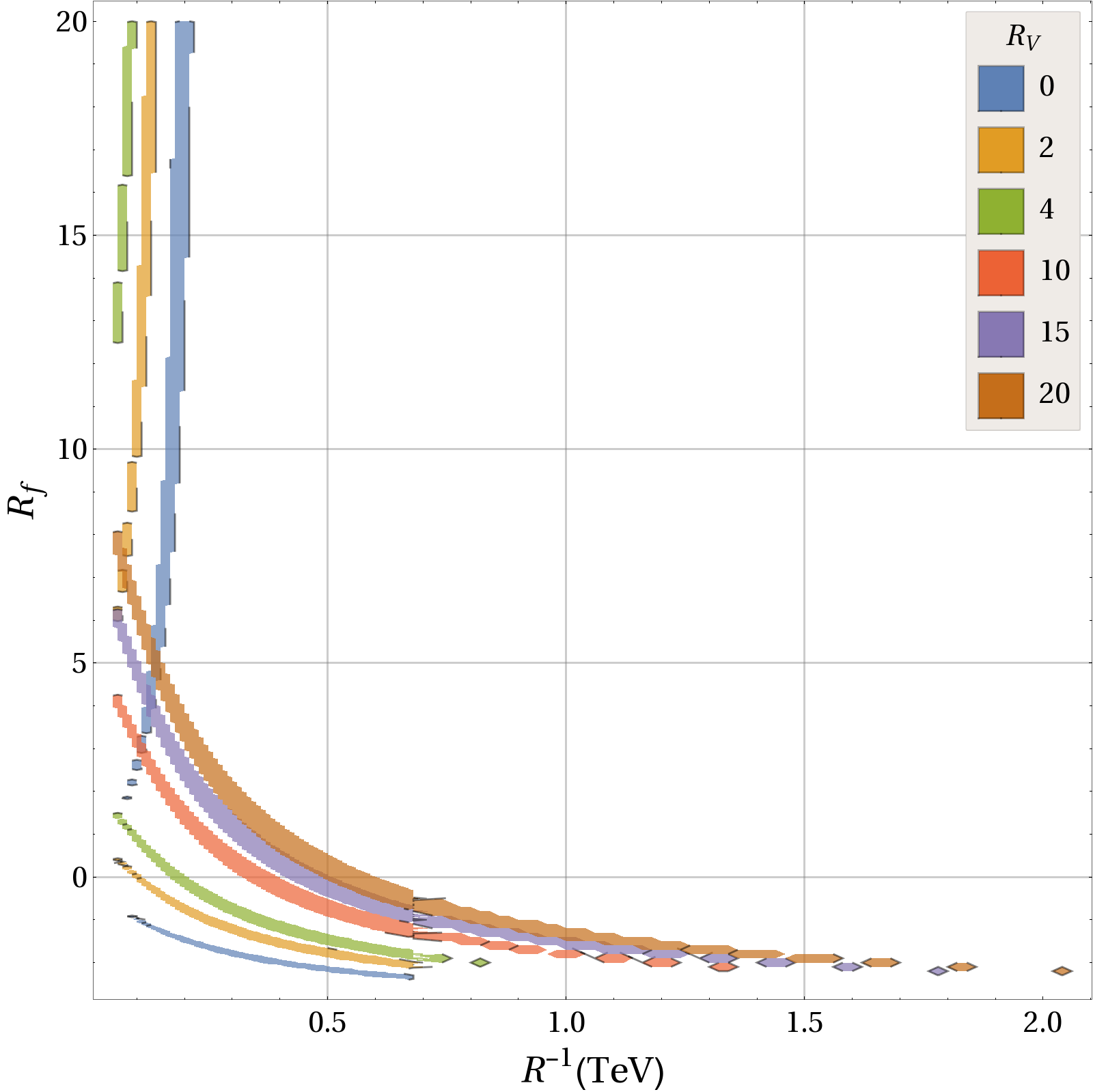}
 \label{fig:rrfdatplt}}
\subfloat[$R_V$ vs. $R^{-1}$]{
\includegraphics[width=5.5cm, height=5.5cm]{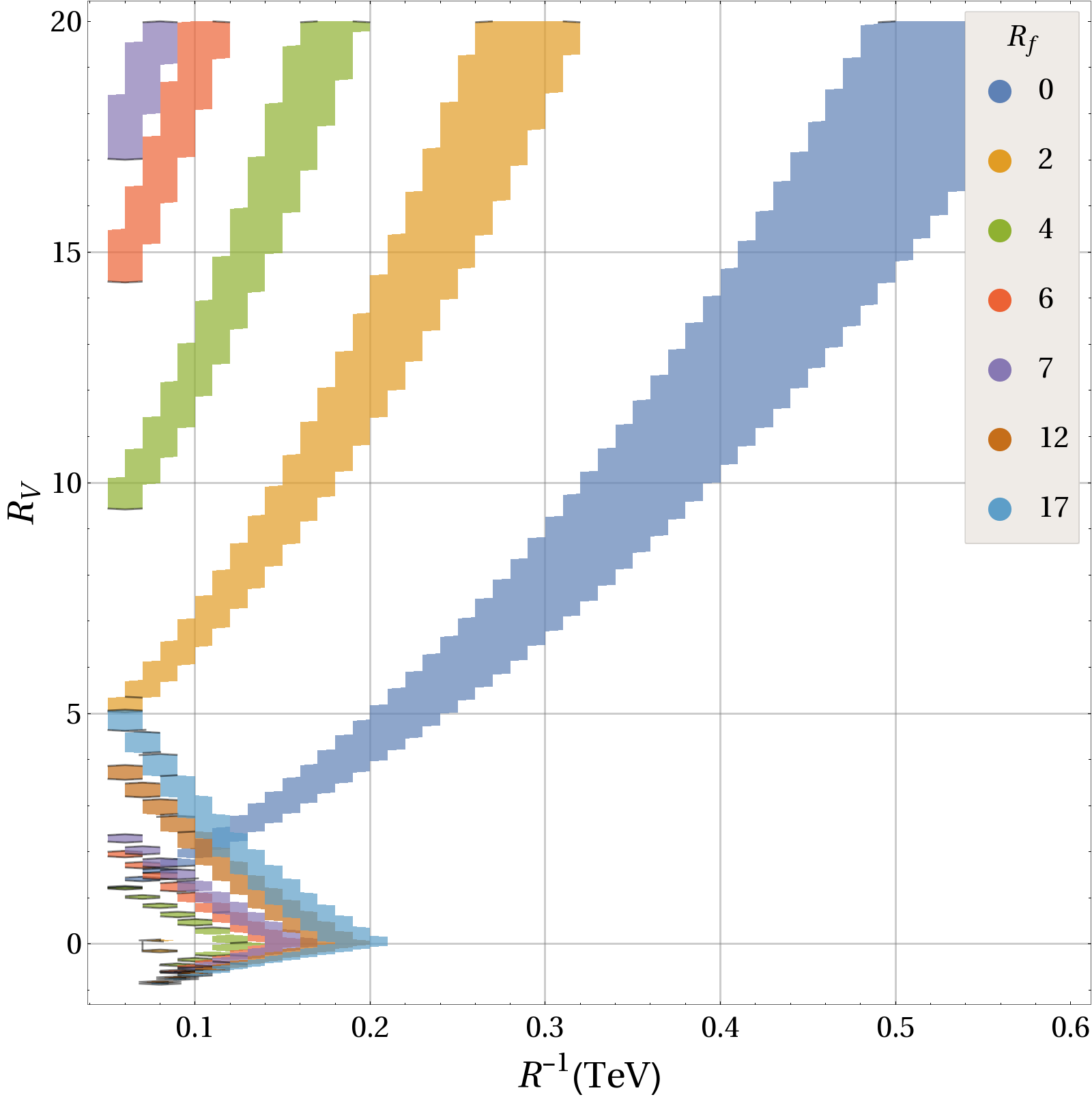}
 \label{fig:rrvdatplt}}
 \caption{Same regions as in fig.~\ref{fig:modpar2}, with multiple distinct values of the $\textquoteleft$other' parameter.}
\label{fig:modpar3}
\end{figure*}

We observe that for all combinations of results shown in fig.~\ref{fig:chiplts}, there is a two-fold ambiguity in the best-fit results. One of these points is closer to SM than the other and as will be clear from the next section, this is the one that is important for us in constraining NMUED. We also note that while the results from Belle and LHCb are consistent with SM within $3 \sigma$ (figs.~\ref{fig:plt2} and \ref{fig:plt5}), for any and all other combination of results, the SM is away from the best fit point by more than $3 \sigma$ in the $C_W$ - $C^{\tau}_{H}$ plane.   

Along with the present theoretical and experimental status, fig.~\ref{rdrdstplt} displays the $1\sigma$ uncertainty ellipse (blue, shaded) corresponding to the fit results with all data in table \ref{tab:res1} (in bold face). It is encouraging that the fit result is completely consistent with the HFAG world average for these ratios as can readily be seen from the figure.

\begin{figure*}\centering
\subfloat[U vs T]{
\includegraphics[height=6.5cm]{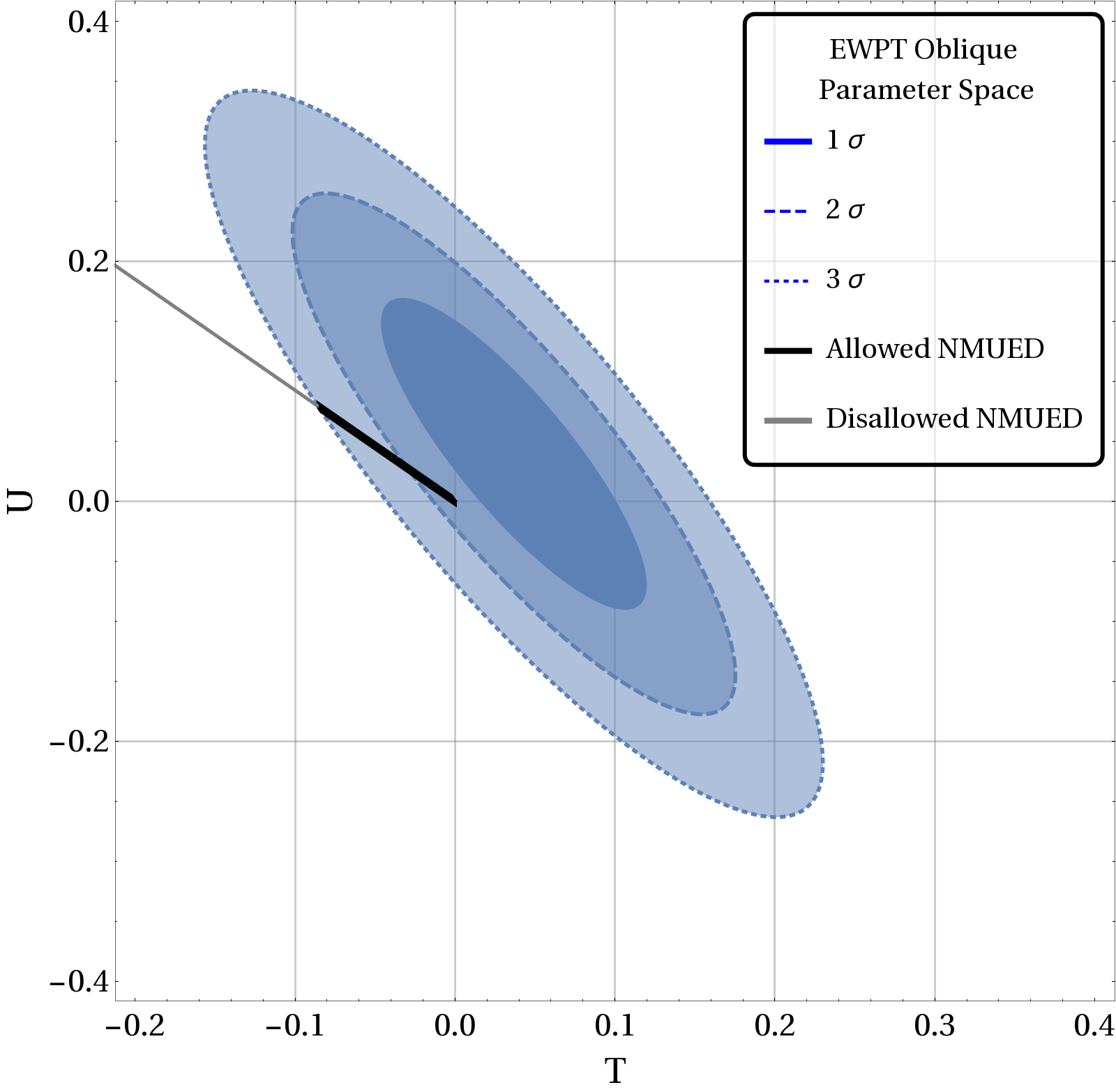}
 \label{fig:TUplt}}
\subfloat[$R_V$ vs $R_f$]{
\includegraphics[height=6.5cm]{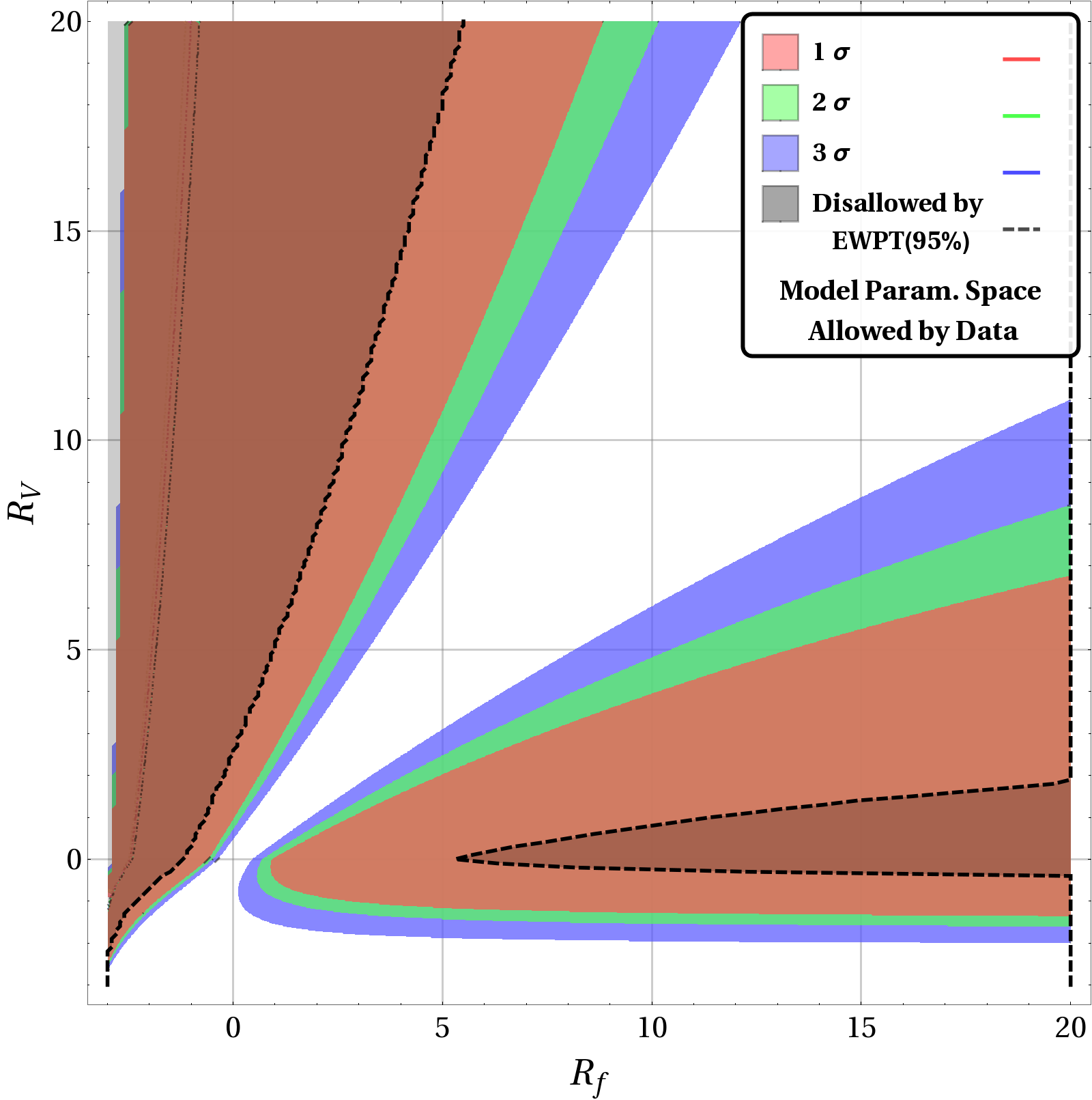}
 \label{fig:rfrvpltparam_EWPT}}
 \caption{EWPT constraints. Fig. (\ref{fig:TUplt}) shows the experimental limits in 1 - 3 $\sigma$ confidence levels. The straight line corresponds to the $U-T$ region populated by NMUED. The black(thick) part of the line is allowed by the experimental limits within $3 \sigma$. Fig. (\ref{fig:rfrvpltparam_EWPT}) shows fig.~\ref{fig:rfrvpltparam} with the parameter space disallowed by EWPT constraints by $2 \sigma$ (dashed contour in fig.~\ref{fig:TUplt}) for all values of $R^{-1}$.}
\label{fig:EWPT}
\end{figure*}

\subsubsection{Constraint from $B^-_c \to \tau^- \bar{\nu}$}

The tauonic decay of $B^-_c$ is linked to $\mathcal{R}(D^{(*)})$ through the same effective general Hamiltonian in eq.~\ref{effH}. The branching fraction for $B_c\rightarrow\tau\nu_{\tau}$ in a particular model (along with  $\mathcal{R}(D^{(*)})$) can hence be used to further constrain the model parameters. The expression for the corresponding branching fraction in the NMUED model is~\cite{Alonso:2016oyd},
\begin{align}
 \nn {\rm Br}(B^-_c \to \tau^- \bar{\nu}) &= \tau_{B^-_c} \frac{m_{B_c} m^2_{\tau} f^2_{B_c} G^2_F \left|V_{cb}\right|}{8 \pi} \left(1 - \frac{m^2_{\tau}}{m^2_{B_c}}\right)^2\\
 & \left|1 + C_W + \frac{m^2_{B_c} m_b}{(m_b + m_c)} C^{\tau}_H\right|^2\,,
\end{align}
where $f_{B_c} = 0.434(15) GeV$ is the $B_c$ decay constant and $\tau_{B^-_c} = 0.507(9) $ps is the $B^-_c$ lifetime.

As is argued in ref. \cite{Alonso:2016oyd}, the $B_c$ lifetime should mainly be accounted for by $b$ and $c$ decays in the $B_c$ meson. As a consequence, only $\lesssim 5\%$ of the measured experimental width, $\Gamma_{B_c} = 1 / \tau_{B_c}$, can be explained by (semi)tauonic modes, including the whole $C_W$ - $C^{\tau}_H$ parameter space that would explain the $\mathcal{R}(D^{(*)})$ excess. Accounting for the maximum possible errors in the calculation , this limit can be relaxed up to $\lesssim 30\%$ of $\Gamma_{B_c}$. 

We thus have chosen an intermediate bound of $\lesssim 10\%$ and in fig.~\ref{pltbc}, we overlay the region excluded by this bound over the main fit result from fig.~\ref{fig:plt1}. As can clearly be seen, this bound completely rules out the $\mathcal{R}(D^{(*)})$ best fit region away from SM, but spares the entire $1 \sigma$ region near SM. Furthermore, as can be seen from fig.~\ref{fig:pltzoom} and is explained in the next section, the entire $C_W$ - $C^{\tau}_H$ parameter space allowed by NMUED is allowed by even the more aggressive bound of $\lesssim 5\%$ of $\Gamma_{B_c}$. Moreover, the allowed region shows that the contribution from the operator involving vector current is mainly responsible for explaining the $\mathcal{R}(D^{(*)})$ excess, which is consistent with the findings of ref. \cite{Bhattacharya:2016zcw}.

\subsection{Model Parameter Estimation}\label{modelparams}

Eqs.~\ref{i1},~\ref{cw} and~\ref{chl} enable one to express the fit parameters $C_W$ and $C_H^\tau$ in terms of the inverse of the radius of compactification ($R^{-1}$) and the scaled BLKT parameters ($R_V(\equiv {r_V}/{R})$, $R_f(\equiv{r_f}/{R})$). The $C_W$ and $C_H^\tau$ fit results (displayed in Table~\ref{tab:res1}) can hence be transformed into constraints on the model parameters. We discuss these constraints using the best fit values of $C_W$ and $C_H^\tau$ in this section and further plot the allowed parameter space in figs.~\ref{fig:modpar1} and \ref{fig:modpar2}. Subsequently, every mention of the allowed parameter space will implicitly assume the $3\sigma$ confidence level unless otherwise stated.

In fig.~\ref{fig:pltzoom}, we show the part of the $C_W$ - $C^{\tau}_H$ space populated by the NMUED parameters. As can clearly be seen, the scalar contribution is orders of magnitude suppressed compared to its gauge boson counterpart. Unlike in other possible similar NP models, this does not mean that $C^{\tau}_H$ is $\approx 0$, due to the reason stated in the penultimate paragraph of the introduction. In fig.~\ref{fig:cwhpltparam}, we zoom in and show the $C_W$ - $C^{\tau}_H$ space allowed by $\mathcal{R}(D^{(*)})$ results, varying all NMUED parameters. As is mentioned in the previous subsection, this whole parameter space is allowed by the constraints coming from $B_c \to \tau \nu$ decay.

We would like to note here that though the BLT parameters can in general be negative or positive, it is evident from eq.~\ref{norm} that for $r_f/R=-\pi$, the zero-mode solution becomes divergent and beyond this limit the fields appear to be ghost like. Therefore, while we show numerical results for negative BLT parameters for the purpose of completeness, any value of BLT parameters lower than $-\pi$ should be discarded. 

It is evident from eq.~\ref{i1} that the overlap integral ($I_n$) vanishes for $R_V=R_f$, making $C_W$ and $C_H^\tau$ vanish as a result\footnote{One should also note that the $C_W$ and $C_H^\tau$ tend to vanish with the increasing values of $R^{-1}$ and asymptotically converges to its SM value as $R^{-1}\rightarrow \infty$}. This corresponds to the SM, which is more than $3 \sigma$ away from the best-fit point according to the experimental results i.e.  we can not explain the excess of the $\mathcal{R}(D^{(*)})$ for $R_V \approx R_f$. In order to do that we need to increase the values of $C_W$ and $C_H^\tau$ which are directly proportional to the overlap integral ($I_n$). This can be done by increasing the absolute value of $|R_f- R_V|$. A higher $|R_f-R_V|$ necessitates a large $R_f$ and a low $R_V$ or vice versa. In fig.~\ref{fig:rfrvpltparam} we show the allowed region of the parameter space obtained from $\mathcal{R}(D^{(*)})$ fit in the $R_f$ vs $R_V$ plane. 
Just to elaborate, the region colored red in that figure corresponds to the red region in fig.~\ref{fig:cwhpltparam} and so on. Due to the reason stated just above, in the limit ($R_V=R_f$), we find a ``discontinuity'' in the allowed parameter space dividing the parameter space in two distinct halves. 

We vary the scaled BLT parameters within their allowed range\footnote{One can obtain the upper limit ($\sim 26$) on the scaled BLT parameters from the unitarity analysis \cite{Jha:2016sre}.} subject to the best fit values of $C_W$ and $C_H^\tau$ dictated by the integrated $\mathcal{\mathcal{R}(D^{(*)})}$ data. Fig~\ref{fig:rrfpltparam}(\ref{fig:rrvpltparam}) shows the variation of $R_f (R_V)$ w.r.t $R^{-1}$ for all possible values of $R_V (R_f)$. Though these plots show the region of parameter space allowed by $\mathcal{R}(D^{(*)})$ data, they do not help us glean any information on the lower bound of $R^{-1}$. Fig.~\ref{fig:modpar3} (where every shown region corresponds to the $1 \sigma$ allowed region of \ref{fig:cwhpltparam}) enables us to put a lower bound on $R^{-1}$ (which we find can reach a value of approximately $1$ TeV). We quote a few benchmark values for the lower limits of $R^{-1}$ for different combinations of $R_f$ and $R_V$ as is evident from figs.~\ref{fig:rrfdatplt} and~\ref{fig:rrvdatplt}:
\begin{itemize}
 \item $R^{-1}$ $\gtrsim 480$ GeV, for $R_f=0,  R_V=15$,
 \item $R^{-1}$ $\gtrsim 550$ GeV, for $R_f=20, R_V=0$ ($R_V=20, R_f=0$),
 \item $R^{-1}$ $\gtrsim 1$ TeV, for $R_f=-1, R_V=20$.
\end{itemize} 
These lower bounds obtained from our analysis is consistent with the studies of $R_b$ \cite{zbb}, branching ratio of  $B_s \rightarrow \mu^+ \mu^-$ \cite{bmm} and $B \rightarrow X_s \gamma$ \cite{bsg}. Fig~\ref{fig:rrfdatplt}(\ref{fig:rrvdatplt}) allows one to extract the maximum value of $R_V$($R_f$) and $R^{-1}$ for a given $R_f$($R_V$).

\subsection{Electroweak precision constraints}\label{ewpt}

We present a brief discussion on the electroweak precision test (EWPT) constraints on the NMUED model in this section. EWPT is an essential and important tool for constraining any form of BSM physics. For the NMUED model, these constraints have been discussed in refs~\cite{flacke, flacke_kong, flacke_Pasold}, albeit in a different approach. In our present work, we follow the approach discussed in~\cite{bmm}.

The corrections to the Peskin-Takeuchi (Oblique) parameters $S$, $T$ and $U$ in NMUED appear through the correction to the Fermi constant, $G_F$ at tree level, which is in stark contrast to the minimal version of the UED model where 
such corrections appear via one loop processes. The corrected Fermi constant $G_F$ can be written as:
\begin{equation}
G_F=G_F^0+\delta G_F,
\end{equation}
where the  0-mode $W^\pm$ exchange contributes to $G_F^0$, while $\delta G_F$ stands for the sum of
the contributions from all non zero (even) $W^\pm$ KK-modes. The effective
Fermi constant can be expressed as
\begin{equation}
G_F^0=\frac{g^2_2}{4\sqrt{2}M^2_W},~~~~~\delta G_F=\sum_{n\geq2}\frac{g^2_2I^2_n}{4\sqrt{2}M^2_{W^{(n)}}},
\end{equation}
where $M^2_{W^{(n)}}$ and $I_n$ are obtained from eqs.~\ref{MWn} and~\ref{i1}.

Following the approach of ref.\;\cite{flacke_kong, flacke_Pasold} the NMUED contributions to the $S$, $T$ and $U$ parameters can be written as:
\begin{eqnarray}
&S_{\rm NMUED}=0,~~T_{\rm NMUED}=-\frac{1}{\alpha}\frac{\delta G_F}{G_F},\nonumber&\\
&U_{\rm NMUED}=\frac{4 \sin^2 \theta_{w}}{\alpha}\frac{\delta G_F}{G_F}.&
\end{eqnarray}
where $g_2$ is the $SU(2)$ gauge coupling constant and $\alpha$ the fine structure constant calculated at $M_Z$. $\theta_w$ is the Weinberg angle.
One can now compare the predictions from NMUED model with the experimental results of $S$, $T$ and $U$, along with their correlations, given in the ref.\;\cite{gfit}, for input Higgs mass $m_h = 125$ GeV and top quark mass $m_t = 173$ GeV.

Fig.~\ref{fig:TUplt} shows the allowed and disallowed ranges for the NMUED model on the U-T plane. We observe that this model is compatible with the EWPT constraints at the $2\sigma$ level.

Fig.~\ref{fig:rfrvpltparam_EWPT} is the result of superimposing these EWPT constraints on fig.~\ref{fig:rfrvpltparam}. We find that the inclusion of the EWPT constraints at $2\sigma$ confidence level results in a reduction of the parameter space allowed by $\mathcal{\mathcal{R}(D^{(*)})}$ for NMUED in the $R_V-R_f$ plane. This also affects the range of allowed lower limits for $R^{-1}$.

\section{Conclusions}

We have investigated the effects of KK-excitations of $W$-boson and charged scalars to the $\mathcal{R}(D^{(*)})$ ratios in a non-minimal Universal Extra Dimensional model in $4 + 1$ dimensions. Here all SM fields can access an extra spatial dimension. This model is characterized by several boundary localized terms (kinetic, Yukawa etc.). These boundary localized terms can be thought of as the counterterms to unknown radiative corrections and the coefficients of these terms can be treated as free parameters of this scenario. Due to the presence of the  these boundary terms the masses and couplings of the KK-excitations have been changed w.r.t the minimal Universal Extra Dimensional model. Using two different types of BLT parameters $R_V$ (for gauge and Higgs sector)  and $R_f$ (fermion sector) along with inverse of the radius of compactification ($R^{-1}$) we have analyzed the $\mathcal{R}(D^{(*)})$ ratios in NMUED model.  

The contributions from the vector gauge bosons and the scalar Higgs bosons have been parametrized in terms of two parameters: $C_W$ which is dimensionless and $C_H^\tau$ which has the dimensions of GeV$^{-2}$. 
In the current analysis of the $\mathcal{R}(D^{(*)})$ ratios, we have neglected masses of the lighter leptons compared to that of $\tau$.
We have performed the fits taking into account several combinations of the available experimental data due to \Babar, Belle and LHCb. We find that the predictions for this model is at par with the HFAG global average for these ratios at $1\sigma$. The best-fit values and errors for $C_W$ and $C_H^{\tau}$ can be translated into constraints for the BLT parameters and $R^{-1}$. For specific values of the BLT parameters ($R_V=20$ and $R_f=-1$), we find that the lower limit of $R^{-1}$ can reach appreciably high values of the order of 1 TeV.
We find that there is a considerable region allowed for these parameters in accordance with the current experimental values for these ratios, as well as constraints coming from $B_c \to \tau \nu$ decay. However, if one considers constraints due to electroweak precision measurements up to $2 \sigma$ level, the allowed parameter space reduces considerably.

\acknowledgments

We thank Prof. Anirban Kundu and Soumitra Nandi for various illuminating comments and discussions.

\end{document}